\documentclass[preprint2]{aastex}

\usepackage{color}
\newcommand{\gps}{\ensuremath{g_{\rm P1}}}
\newcommand{\rps}{\ensuremath{r_{\rm P1}}}
\newcommand{\ips}{\ensuremath{i_{\rm P1}}}

\defcitealias{K18a}{K18a}

\slugcomment{Accepted for publication in ApJ}

\shorttitle{M31 PAndromeda Cepheid sample observed in four HST bands}
\shortauthors{Kodric et al.}

\begin{document}

\title{M31 PAndromeda Cepheid sample observed in four HST bands}

\author{Mihael Kodric\altaffilmark{1,2}, Arno Riffeser\altaffilmark{1,2}, Stella Seitz\altaffilmark{1,2}, Ulrich Hopp\altaffilmark{1,2}, Jan Snigula\altaffilmark{2,1}, Claus Goessl\altaffilmark{1,2}, Johannes Koppenhoefer\altaffilmark{2,1}, Ralf Bender\altaffilmark{2,1} }

\email{kodric@usm.lmu.de} 

\altaffiltext{1}{University Observatory Munich, Scheinerstrasse 1, 81679 Munich, Germany}
\altaffiltext{2}{Max Planck Institute for Extraterrestrial Physics, Giessenbachstrasse, 85748 Garching, Germany}

\begin{abstract}
Using the M31 PAndromeda Cepheid sample and the HST PHAT data we obtain the largest Cepheid sample in M31 with HST data in four bands. For our analysis we consider three samples: A very homogeneous sample of Cepheids based on the PAndromeda data, the mean magnitude corrected PAndromeda sample and a sample complementing the PAndromeda sample with Cepheids from literature. The latter results in the largest catalog with 522 fundamental mode (FM) Cepheids and 102 first overtone (FO) Cepheids with F160W and F110W data and 559 FM Cepheids and 111 FO Cepheids with F814W and F475W data. The obtained dispersion of the Period-Luminosity relations (PLRs) is very small (e.g. 0.138 mag in the F160W sample I PLR). We find no broken slope in the PLRs when analyzing our entire sample, but we do identify a subsample of Cepheids that causes the broken slope. However, this effect only shows when the number of this Cepheid type makes up a significant fraction of the total sample. We also analyze the sample selection effect on the Hubble constant.
\end{abstract}

\keywords{catalogs -- cosmology: distance scale -- galaxies: individual(M31) -- Local Group -- stars: variables: Cepheids}

\section{Introduction}

Cepheids follow a Period-Luminosity relation (PLR), sometimes also referred to as Leavitt Law (\citet{1908AnHar..60...87L} and \citet{1912HarCi.173....1L}), making them standard candles used to determine extragalactic distances. They are also an essential part of the cosmic distance ladder used to determine the Hubble constant ($\mathrm{H}_0$, \citet{1929PNAS...15..168H}). \citet{2016ApJ...826...56R} determine $\mathrm{H}_0$ with an uncertainty level of only 2.4\% (see \citet{2010ARA&A..48..673F} for a summary of previous projects determining $\mathrm{H}_0$). The treatment of outliers in the PLR and therefore the uncertainty level in $\mathrm{H}_0$ are still under debate (see e.g. \citet{2014MNRAS.440.1138E} and \citet{2015arXiv150707523B}). There is also an ongoing debate in the literature if the PLR has a broken slope \citep{2009A&A...493..471S}. \citet{2008A&A...477..621N} find a broken slope, but linear relations in the Ks band and the Wesenheit PLR. \citet{2013ApJ...764...84I} however find linear Wesenheit PLRs, while \citet{2013MNRAS.431.2278G} find no linear relations but rather exponential ones. In \citet{K15} we find a significant broken slope in all relations. Also the effect of metallicity  is debated in the literature (see e.g. \citet{2011ApJ...734...46F} and \citet{2011ApJ...741L..36M}).
Distance determinations with Cepheids need anchor galaxies with well determined distances. Some widely used anchor galaxies are: the Milky Way (see e.g. \citet{2016ApJ...825...11C} for an effort to obtain parallaxes with HST and \citet{2017EPJWC.15202003C} for the Gaia project), the Magellanic clouds (see e.g. \citet{2015AcA....65....1U} and \citet{2015AcA....65..297S} for the OGLE survey and \citet{2015ApJ...815...28G} for the Araucaria project), M106 (e.g. \citet{2013ApJ...775...13H} and \citet{2015AJ....149..183H}) and M31 (see e.g. \citet{2006A&A...459..321V}, \citet{2012ApJ...745..156R} (hereafter R12), \citet{K13} (hereafter K13), \citet{K15} (hereafter K15), \citet{2015MNRAS.451..724W} and \citet{K18a} (hereafter K18a).

PAndromeda \citep{2012AJ....143...89L}, a key project of Panstarrs \citep{2016arXiv161205560C}, has found 2639 Cepheids in M31 \citepalias{K18a}. We use this sample in combination with PHAT \citep{2012ApJS..200...18D} HST photometry to obtain the largest Cepheid sample in M31 with HST photometry in four bands. For more information about M31 as a distance anchor we refer to \citetalias{K18a} and \citet{2017arXiv170102507L}.

This paper is structured as follows: In section \ref{DR} we discuss the data reduction and the different Cepheid samples (section \ref{samples}) we analyze. We also discuss our mean magnitude correction (section \ref{correction}) and outlier rejection (section \ref{outlier}). In section \ref{results} we discuss the results and follow it with a conclusion in section \ref{conclusion}.

\section{Data reduction\label{DR}}

We combine ground-based observations of Cepheids with HST data from M31. The PHAT survey \citep{2012ApJS..200...18D}, that covers about a third of the disk of M31 in six filters (F275W, F336W, F475W, F814W, F110W and F160W), is used to obtain photometry of the Cepheids, while the ground based data where used to detect the Cepheids and further provides the period and location of the objects.

The PHAT data was retrieved from the MAST archive and the photometry is obtained with DOLPHOT \citep{2000PASP..112.1383D}.
We use the pipeline we describe in K15 to identify the Cepheid in the HST data. The pipeline automatically aligns the HST images to the K18a PS1 reference frame and produces  difference frames that are used to determine the most likely position of the Cepheid. As in K15 we visually inspect the result of the pipeline, i.e. the stamp outs and difference frames in all available HST bands and the corresponding PS1 reference frames. We modified the pipeline to also provide variability maps from PS1 and color magnitude diagrams with the corresponding instability strip as defined in K18a. The variability maps show how often each pixel has an one sigma deviation over the background in all difference frames normalized to the total number of epochs available in the pixel. The color magnitude diagrams are constructed per HST instrument, i.e. F275W-F336W vs. F336W for WFC3/UVIS, F475W-F814W vs. F814W for ACS/WFC and F110W-F160W vs. F160W for WFC3/IR. The combination of all this information allows us to identify the Cepheid among the HST sources. Note that we do not require the source to be inside the edges of the instability strip we defined in K18a, since we work with random phased magnitudes in the identification process. Nevertheless the combination of the location in the color magnitude diagram with the other available information, e.g. the variability map makes the correct identification very robust, i.e. the Cepheid can be identified unambiguously. 

While the WFC3/UVIS filters are useful to identify the correct source, in this paper we look only at ACS/WFC and WFC3/IR filters.
We abbreviate those filters as g (F475W), I (F814W), J (F110W) and H (F160W). Note that we do not perform a filter conversion, this is only done for better readability in tables and figures. Throughout the paper we use the Vega magnitude system.

DOLPHOT detects the sources in one filter and performs the photometry on those detected positions in both filter pairs (i.e. in the two filters corresponding to the same instrument). We run DOLPHOT twice so that the source detection is done in each filter. This means that we have two measurements for each source, but we only use the measurement were the source detection was done in the same filter. In some cases the pipeline fails to automatically align a HST frame in a certain filter to the PS1 reference frame. In this case we use the measurement where the source detection was done in the other filter, where the alignment did work. This could introduce a systematic effect if there is a close source to the Cepheid that is not detected in the filter where the source detection was done and therefore has not been subtracted, but would have been detected in the other filter. We do not see such systematic effect on the PLR, but we flag those Cepheids in the published tables.

We define the following, reddening-free Wesenheit magnitudes:
\begin{eqnarray}
	W_{gI} = g - 1.876 \cdot (g-I) \label{eqn_W_gI} \\
	W_{JH} = J - 2.388 \cdot (J-H) \label{eqn_W_JH}
\end{eqnarray}
where the Wesenheit index is obtained from \citet{2011ApJ...737..103S} with an extinction law of $R_V=3.1$.

We also perform an extinction correction and use these corrected magnitudes throughout this paper. The extinction coefficients are obtained from \citet{2011ApJ...737..103S} using an extinction law of $R_V=3.1$. The $\mathrm{E(B-V)}$ map from \citet{2009A&A...507..283M} is used to correct for the extinction in M31. Since we do not know whether the Cepheid has the total amount of dust that \citet{2009A&A...507..283M} predicts for this location or if it has no dust from M31 in front of it, we correct with half the color excess given for that location. We also correct for the foreground extinction by using the value determined by  
\citet{1998ApJ...500..525S} ($\mathrm{E_{fg}(B-V)=0.062}$) and apply the correction factor 0.86 for the foreground determined by \citet{2011ApJ...737..103S}.

\subsection{Cepheid samples\label{samples}}

As described in the previous section we use ground based data to determine the Cepheid location, period and type.
Based on which data is used, we distinguish between sample I, sample II and sample III.

\underline{Sample I} is based on the K18a sample. In K18a we obtain the largest M31 Cepheid sample using the Pan-STARRS1 PAndromeda data in the \gps, \rps~and \ips~band. From the 2639 Cepheids in K18a we find 824 in the PHAT data. 780 have WFC3/IR data (i.e. J and H band data) and 821 have ACS/WFC data (i.e. g and I band data). Based on the K18a type classification of the 824 Cepheids 171 are unclassified (UN) Cepheids, 486 fundamental mode (FM) Cepheids, 92 first overtone (FO) Cepheids and 75 type II (T2) Cepheids. In K15 where we made a similar analysis as in this paper but based on the K13 data (K18a is based on K13, same as this paper is based on K15) we identified 492 Cepheids with J and H band data, making the sample in this paper a factor of $\sim1.6$ larger. Of the 780 Cepheids with NIR data 53 have more than one epoch and 755 have more than one epoch from the 821 Cepheids with g and I band data. We use the mean magnitude in the cases where we have more than one epoch. A weighted mean magnitude would bias the measurement towards brighter magnitudes since the error is smaller for brighter magnitudes. This means that sample I is random phased.

\underline{Sample II} is based on multiple Cepheid catalogs. We start with the largest available catalog which is K18a with 2639 Cepheids (this means that sample I is a subsample of sample II) and add the Cepheids from the 2009 K13 Cepheids that are not present in K18a. Next we add the 416 Cepheids from \citet{2006A&A...459..321V} that are not within one arcsecond of the combined catalog. The 332 Cepheids from the DIRECT survey (\citet{1998AJ....115.1894S}, \citet{1998AJ....115.1016K}, \citet{1999AJ....117.2810S}, \citet{1999AJ....118..346K}, \citet{1999AJ....118.2211M}, \citet{2003AJ....126..175B}), the 126 Cepheids from the WECAPP survey (\citet{2006A&A...445..423F}) and the 68 Cepheids from \citet{2012ApJ...745..156R} are added next in the same manner. 1179 Cepheids have PHAT data, 824 from K18a, 148 from K13, 79 from \citet{2006A&A...459..321V}, 54 from DIRECT, 69 from WECAPP and 5 from \citet{2012ApJ...745..156R}. The Cepheid types are assigned according to the catalogs in the case of K18a and K13. For WECAPP Cepheids DC corresponds to FM and W to T2 and other types we classify as UN. The \citet{2006A&A...459..321V} Cepheids are classified as FM or FO based on private communication. All \citet{2012ApJ...745..156R} Cepheids are FM and the DIRECT Cepheids do not have a type assigned so we classify them as UN. Sample II consists of 298 UN Cepheids, 604 FM Cepheids, 122 FO Cepheids and 155 T2 Cepheids.
1113 have J and H band data and 1176 have g and I band data. For 45 Cepheids in sample II we were not able to determine which source is the Cepheid. This might be due to uncertainties in the Cepheid coordinates. We flag those Cepheids and do not use them. Sample II is also random phased and we also use the mean magnitude were multiple epochs are available. Sample II is a factor of $\sim 2.4$ larger than the sample in K15.

\begin{figure*}
  \centering
  \includegraphics[width=0.9\linewidth]{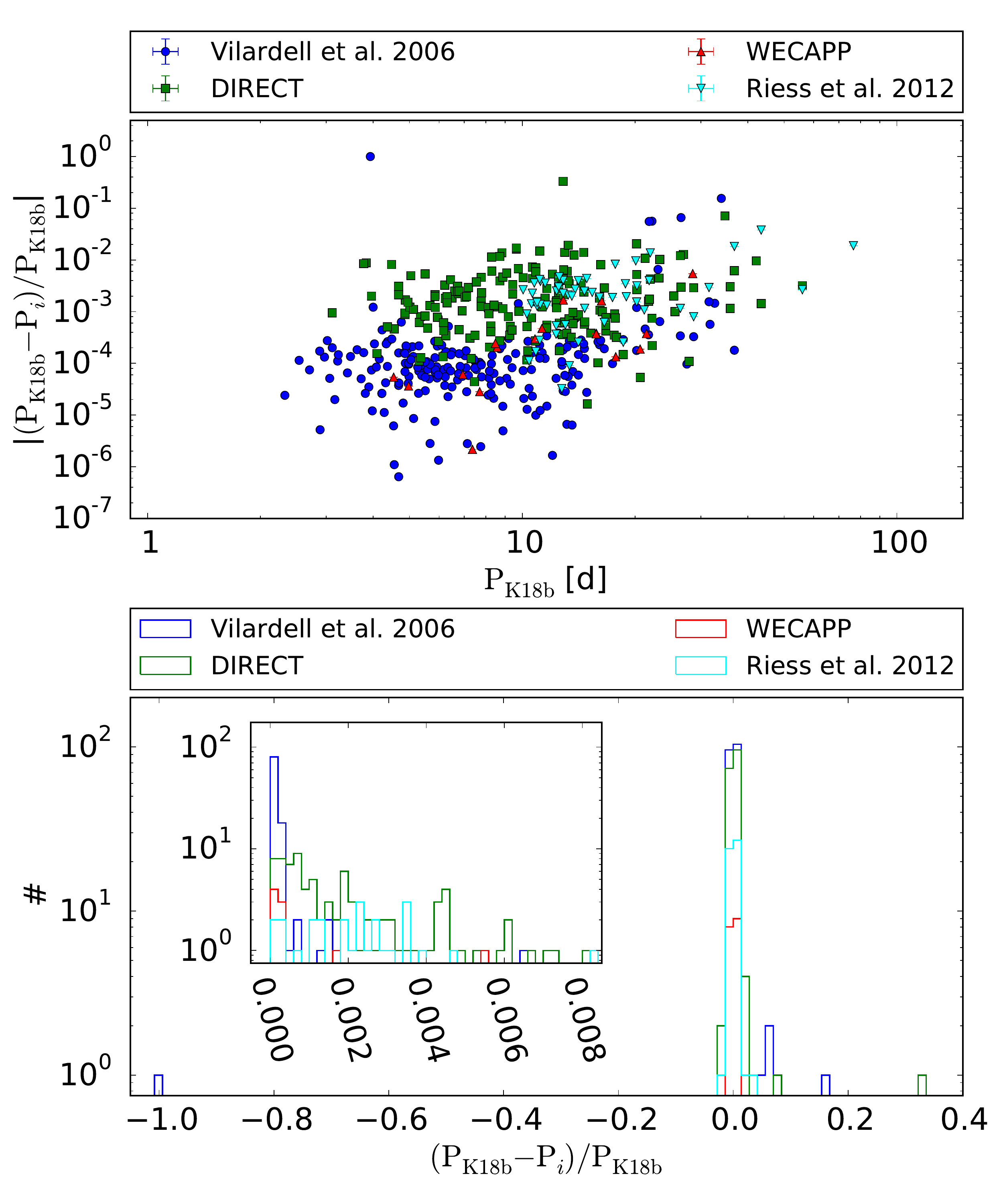}
  \caption{Period comparison of K18a Cepheids that are in K18b with literature Cepheids. Due to the 420 observed epochs over four years the PAndromeda periods are very precise. The top panel shows that the relative period difference is small, but increases towards longer periods. This is due to the fact that the baseline that is used to determine the literature periods is not long enough to determine long periods as accurately as in the PAndromeda survey. The bottom panel shows that the periods scatter symmetrically around the PAndromeda periods, which means that there will be no systematic effect introduced into the sample II PLR from the literature periods.    
  \label{fig_per_diff}}
\end{figure*}

The Cepheid type is determined by only one method in sample I in contrast to sample II. This is also true for the selection criteria used to determine if a variable is a Cepheid, which makes sample I more homogeneous. The PAndromeda Cepheid sample in K18a has up to 420 epochs which makes it not only the largest Cepheid sample in M31 but also the sample with which has the most precise periods\footnote{See the appendix in K18a for the period difference of long period Cepheids for the case where 1 or 4 seasons of M31 data were used to obtain the periods.}. Therefore the periods are more accurate in sample I than in sample II, but sample II has more Cepheids. In order to check how well the periods from the literature compare to the K18a periods we matched the different samples with a matching-radius of one arcsecond (this means that those are the literature Cepheids that are not included in sample II since they are already in K18a). As can be seen in figure \ref{fig_per_diff} the relative period difference is small, but increases towards longer periods. This is because the baseline in most literature samples is not long enough to determine long periods accurately. The exception is the WECAPP survey which has a baseline of three years which is comparable to the four years of PAndromeda, but the WECAPP surveys period determination suffers from the smaller telescope size (80cm) and consequently poorer photometry (also WECAPP covered mostly the center of M31 where crowding strongly influences the photometry). Figure \ref{fig_per_diff} also shows that apart from a few outliers and one aliasing case the periods scatter symmetrically around the K18a periods, which means that the sample II PLR will not have a systematic effect introduced due to the literature periods. Note that in order to draw this conclusion we have to assume that the periods of the literature Cepheids that are included in sample II behave the same as those shown in figure \ref{fig_per_diff}.   

\underline{Sample III} consists of the FM Cepheids from sample I. In contrast to sample I we perform a mean magnitude correction so that this sample is not random phased. The correction is described in the next section.

\subsection{Mean magnitude correction\label{correction}}

In order to determine mean magnitudes for sample III we make use the optical light curves obtained in K18a in the Pan-STARRS1 \gps, \rps~and \ips bands. The idea is to transform the optical light curve $m_{\mathrm{K18a}}(t)$ into a light curve belonging to the observed HST filters (g, I, J and H). The reason sample III is based on sample I is that the random phased PHAT measurements were taken during the K18a observations. This means that the period determined in K18a is also very precise for the PHAT measurements. During the baseline of observations the period is precise and therefore the phase of the light curve can be determined. For times outside this baseline even small uncertainties in the period can add up so that the prediction of the phase is completely wrong (e.g. maximum light is expected but in fact minimum light would be observed).

In order to transform the \rps~band light curve into e.g. the H band light curve we need to determine three parameters, i.e. the amplitude ratio $A_{H,r}$, the phase lag $\Theta_{H,r}$ and an offset $s_{H,r}$. The phase lag between the \gps~and \rps~and \ips~band has been determined as zero in K18a. Therefore we are free to choose which band from K18a we want to base the mean magnitude correction on. Since the \rps~band has the most epochs we use the \rps~band light curves. \cite{2015A&A...576A..30I} determined the phase lag between the V and J band and between V and H band. Since the V band is between the \gps~and \rps~band and the phase lag between those is zero the \cite{2015A&A...576A..30I} relations in their figure 3 are also valid for the \rps~band and not only for the V band.
The amplitude ratio between two filters is assumed to be the same for all Cepheids (see K18a), while the offset will be different for each Cepheid due to crowding in the optical band and differences in reddening/extinction from object to object that might not have been correctly accounted for in our extinction correction. Therefore we use the Cepheids with at least two measurements to determine the amplitude ratio. 446 of the 484 Cepheids in the g and I bands fulfill this requirement, while only 36 of the 462 Cepheids in the J and H bands have at least two PHAT measurements. In order to obtain the amplitude ratio $A_{H,r}$ we solve 
\begin{eqnarray}
\chi^2 = \sum_{i,j} {\left( \frac{m_{H,i,j} - s_{H,r,i} - A_{H,r} \cdot m_{r,i,j}}{\sigma_{i,j}} \right)}^2
\end{eqnarray}
with
\begin{eqnarray}
m_{r,i,j} \equiv m_r(P_{i}\cdot\Theta_{H,r}(P_{i})+t_{i,j})
\end{eqnarray}
where the index i denotes the different Cepheids and the index j the measurements of this Cepheid. $m_{H,i,j}$ are the PHAT measurements, $m_r$ the r band magnitudes at the times $t_{i,j}$ where the PHAT measurements were taken and corrected by the phase lag (that depends on the Period $P$ according to \cite{2015A&A...576A..30I}). Note that the offset is different for each Cepheid while the amplitude ratio is the same. The two parameters are degenerate and therefore it is not possible to fit the amplitude ratio individually for each Cepheid. We obtain for the amplitude ratios
\begin{eqnarray}
A_{g,r} = \frac{A_g}{A_r} = 1.3259 \pm 0.0008 \\
A_{I,r} = \frac{A_I}{A_r} = 0.7345 \pm 0.0005 \\
A_{J,r} = \frac{A_J}{A_r} = 0.624 \pm 0.002 \\
A_{H,r} = \frac{A_H}{A_r} = 0.498 \pm 0.002 
\end{eqnarray}
where the errors are determined from the $\chi^2$ minimization.
As expected the amplitude ratio gets smaller for longer wavelengths \citep{1991PASP..103..933M}. In a last step we use these ratios and determine the offset to each Cepheid. The resulting light curve is used to determine the mean magnitude.

This approach to obtain a mean magnitude from random phased observations works reasonably well, but there are some problems. For most Cepheids with more than one measurement the obtained light curve fits the measurements well. Of course if there is only one measurement it will fit the light curve perfectly (since there is one data point for one free parameter). But there are also cases where e.g. the measurements would require the light curve to decline but the obtained light curve is rising. In these cases the phase lag seems to be wrong. This is not surprising since the phase lag relation has a non negligible dispersion (see \cite{2005PASP..117..823S}, \cite{2015A&A...576A..30I} and K18a). The determined amplitude ratios might also be biased by Cepheids that are very crowded in the ground based r band. While the mean absolute magnitude difference between the corrected mean magnitude and the magnitude from averaging the random phased measurements ranges from up to 0.16 mag in the g band down to 0.08 mag in the H band, the mean magnitude difference ranges from up to 0.0276 mag in the g band down to 0.0008 mag in the J band. This means that while each individual Cepheids magnitude changes due to the correction, the magnitudes on average stay the same. As will be discussed later the dispersions in the PLRs do not improve significantly (in the case of the $W_{JH}$ PLR the dispersion even increases). Nevertheless this mean magnitude correction is still useful to check if there are systematics in the not corrected PLR.

\begin{figure*}
\centering
\includegraphics[width=0.9\linewidth]{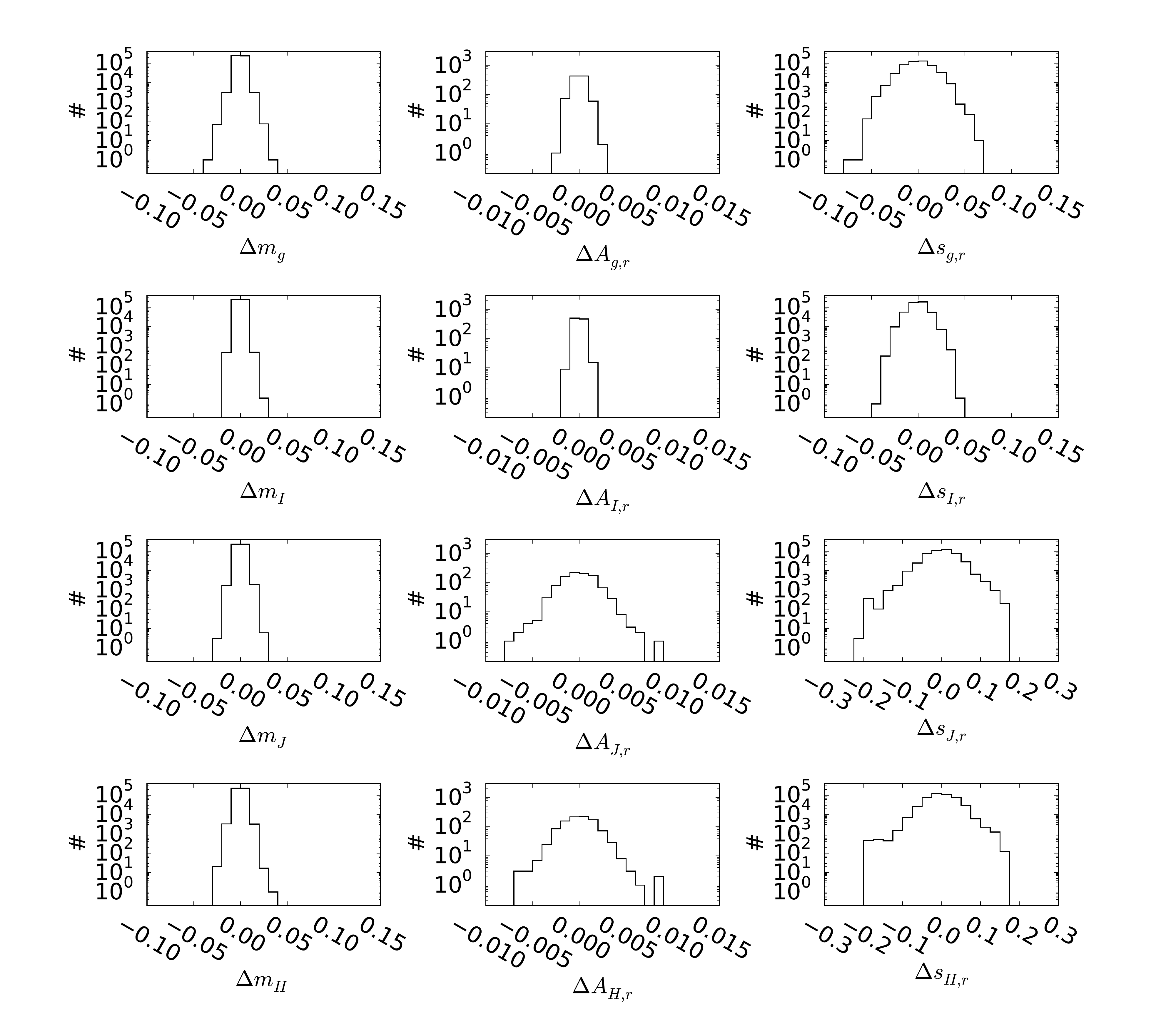}
\caption{Error estimates from a simulation of the mean magnitude correction procedure described in section \ref{correction}. The differences between the value obtained from the simulation and the model are shown as histograms. The errors in the mean magnitude ($\Delta m_i$) are smaller than the values we add quadratically to the error of each Cepheid, while the errors in the amplitude ratios ($\Delta A_{i,r}$) are larger than the errors obtained from the $\chi^2$ minimization. The distribution of the offset ($\Delta s_{i,r}$) are also shown.
\label{fig_simul}}
\end{figure*}

The dispersion between the PHAT measurements and the obtained light curve over all Cepheids is added quadratically to the error of each Cepheid (0.109 mag in g, 0.046 mag in I, 0.036 mag in J and 0.041 mag in H). In order to additionally estimate the errors of the mean magnitude correction we perform a simulation of our procedure. For each Cepheid we start with its K18a \rps~band light curve and transform it to the g, I, J and H band using an amplitude ratio and an offset. We chose constant amplitude ratios close to the ones shown in the equations above and constant offsets that are close to the mean offsets determined from the PHAT data. From these analytic light curves we randomly draw as many points as there are PHAT observations. This means that if there are two PHAT epochs in e.g. the J band, our simulated light curve also has two random epochs in the J band. Each magnitude in the simulated light curve is randomly drawn from a Gaussian that has its mean at the magnitude of the simulated light curve at the drawn epoch and the PHAT magnitude error as its standard deviation. So the magnitudes of our simulated PHAT light curve will scatter around the model light curve. We than perform the mean magnitude correction described above, i.e. we fit the amplitude ratio simultaneously for all Cepheids with two or more epochs. For each Cepheid we know the mean magnitude from the simulated light curve and the amplitude ratio and the offset used to obtain that light curve. We perform 1000 iterations of this procedure. This means that we have 1000 measurements for the amplitude ratio and 486000 measurements for the mean magnitude and offset. In Figure \ref{fig_simul} we show the difference in mean magnitude, the difference in amplitude ratio and the difference in the offset. As can be seen the errors in the mean magnitude are smaller than the values we add quadratically to the error of each Cepheid, while the errors in the amplitude ratios are larger than given in the equations above. 

\subsection{Outlier rejection\label{outlier}}

\begin{figure*}
\centering
\includegraphics[width=0.9\linewidth]{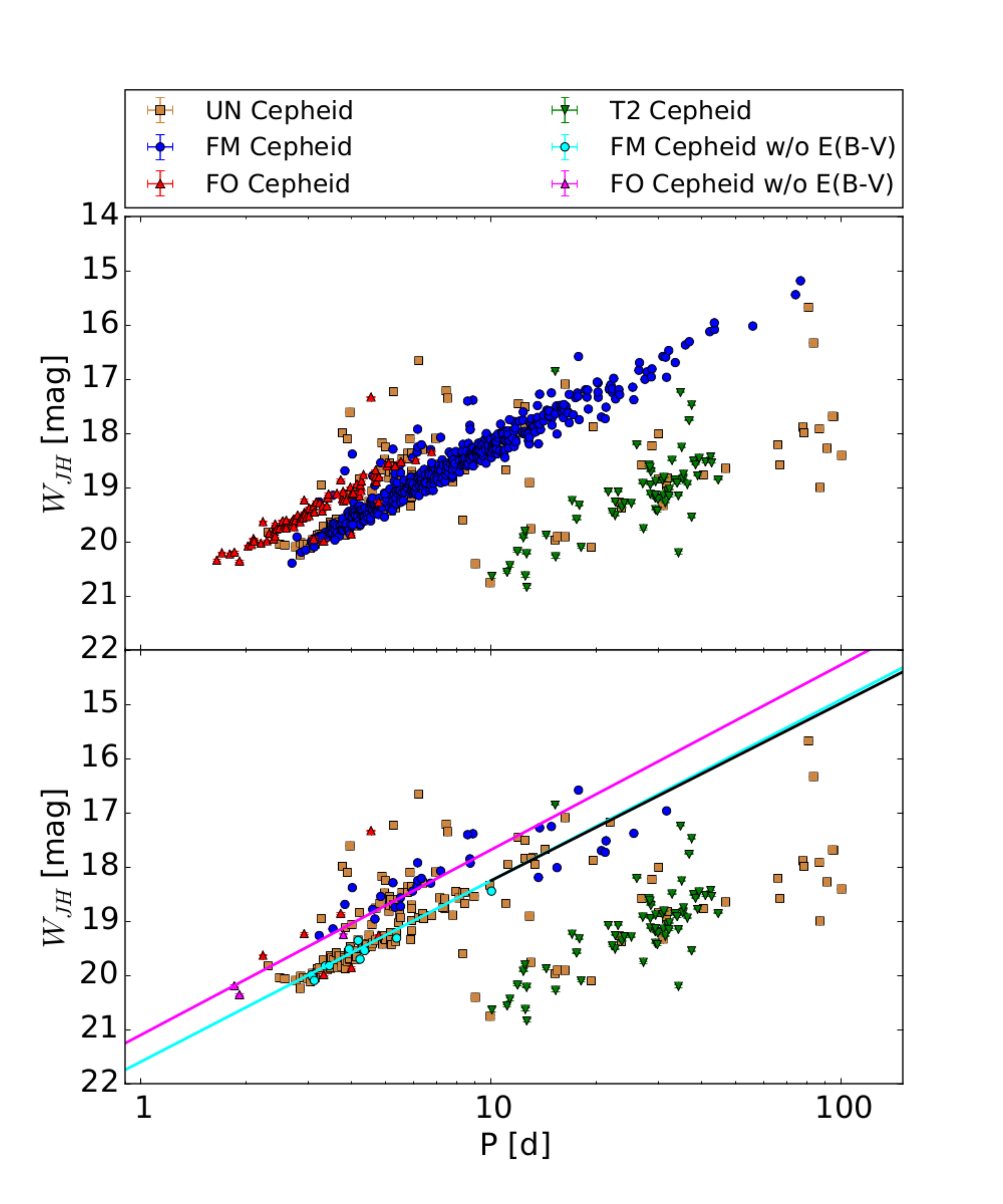}
\caption{Sample I $\mathrm{W}_{\mathrm{JH}}$ Period-Luminosity relation (PLR) \textit{Top panel}: Complete sample before the clipping (462 fundamental mode (FM), 87 first overtone (FO), 67 type II (T2) and 164 unclassified (UN) Cepheids)
\textit{Bottom panel}: Cepheids that are clipped (32 FM and 8 FO Cepheids) or rejected because there is no color excess (12 FM and 3 FO Cepheids) or are not used in this paper (164 UN and 67 T2 Cepheids). The PLRs shown as different lines correspond to those of figure \ref{fig_W_JH}. There are very few outliers and the clipped Cepheids are mostly excluded because the Cepheid type was misclassified based on the (ground-based) K18a data. The UN Cepheids that are excluded follow the PLRs, so excluding them does not bias the sample.
\label{fig_clipping-sample-I-W_JH}}
\end{figure*}

\begin{figure*}
\centering
\includegraphics[width=0.9\linewidth]{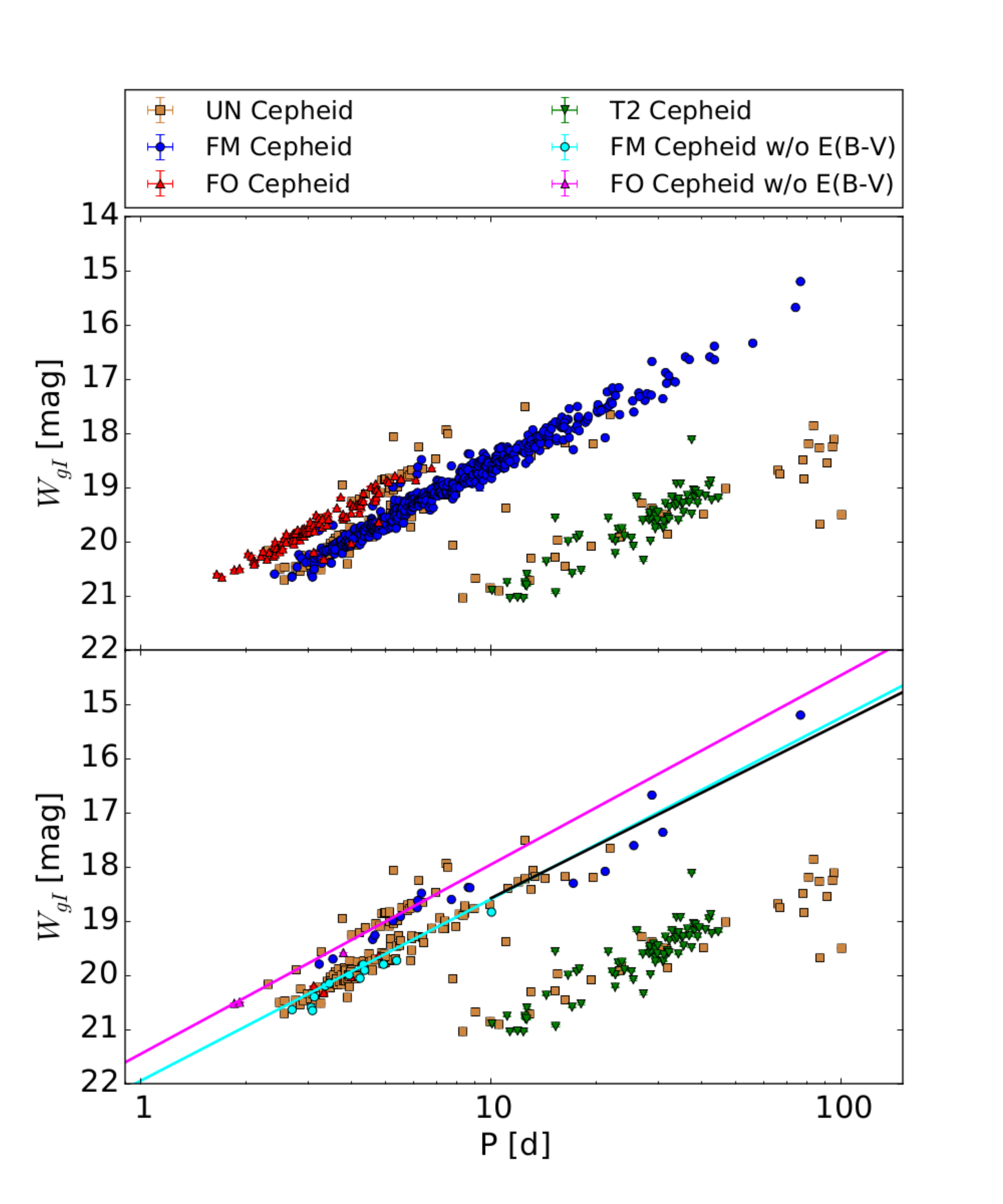}
\caption{Sample I $\mathrm{W}_{\mathrm{gI}}$ PLR \textit{Top panel}: Complete sample before the clipping (484 FM, 91 FO, 75 T2 and 171 UN Cepheids)
\textit{Bottom panel}: Cepheids that are clipped (18 FM and 4 FO Cepheids) or rejected because there is no color excess (13 FM and 3 FO Cepheids) or are not used in this paper (171 UN and 75 T2 Cepheids). The PLRs shown as different lines correspond to those of figure \ref{fig_W_gI}. Same as in figure \ref{fig_clipping-sample-I-W_JH} the UN Cepheids do not bias the sample and the clipped Cepheids are mostly due to type missclassification. There are even fewer outliers than in the $\mathrm{W}_{\mathrm{JH}}$ PLR, which shows how extremely well the Cepheid detection works in the optical bands in K18a. 
\label{fig_clipping-sample-I-W_gI}}
\end{figure*}

\begin{figure*}
\centering
\includegraphics[width=0.9\linewidth]{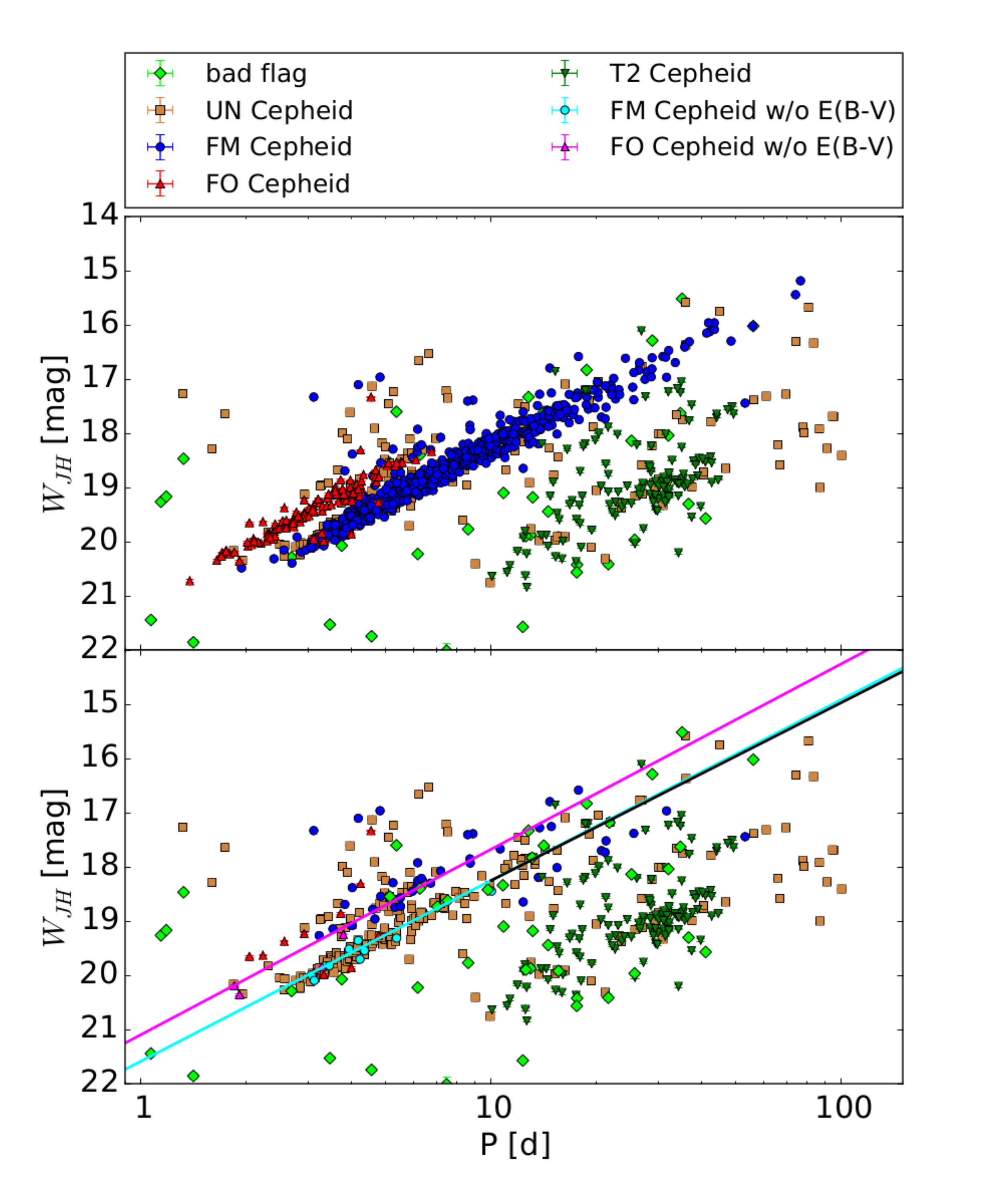}
\caption{Sample II $\mathrm{W}_{\mathrm{JH}}$ PLR \textit{Top panel}: Complete sample before the clipping (574 FM, 117 FO, 124 T2, 255 UN Cepheids and 43 Cepheids with a bad quality flag) \textit{Bottom panel}: Cepheids that are clipped (40 FM and 11 FO Cepheids) or rejected because there is no color excess (12 FM and 4 FO Cepheids) or are not used in this paper (255 UN and 124 T2 Cepheids) or have a bad quality flag (43 Cepheids). Note that some of the flagged Cepheids are outside the region shown here. By adding literature Cepheids to sample II the outlier rejection gets more important compared to sample I shown in figure \ref{fig_clipping-sample-I-W_JH}.
\label{fig_clipping-sample-II-W_JH}}
\end{figure*}

\begin{figure*}
\centering	
\includegraphics[width=0.9\linewidth]{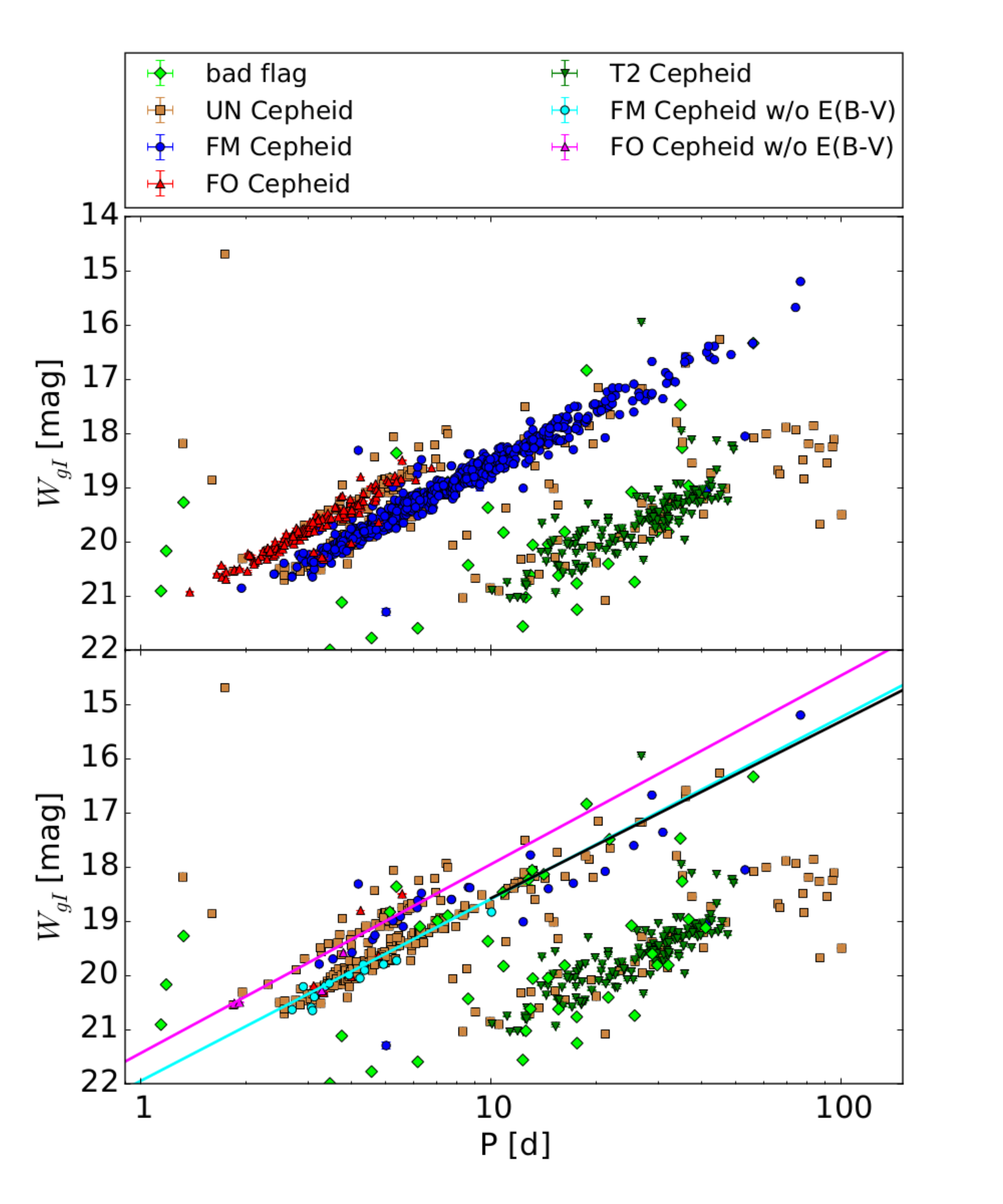}
\caption{Sample II $\mathrm{W}_{\mathrm{gI}}$ PLR \textit{Top panel}: Complete sample before the clipping (601 FM, 121 FO, 141 T2, 268 UN Cepheids and 45 Cepheids with a bad quality flag) \textit{Bottom panel}: Cepheids that are clipped (28 FM and 6 FO Cepheids) or rejected because there is no color excess (14 FM and 4 FO Cepheids) or are not used in this paper (268 UN and 141 T2 Cepheids) or have a bad quality flag (45 Cepheids). Note that some of the flagged Cepheids are outside the region shown here. Compared to the $\mathrm{W}_{\mathrm{JH}}$ PLR shown in figure \ref{fig_clipping-sample-II-W_JH} there are less outliers but again the outlier rejection gets more important compared to sample I shown in figure \ref{fig_clipping-sample-I-W_gI}.
\label{fig_clipping-sample-II-W_gI}}
\end{figure*}

\begin{figure*}
\centering
\includegraphics[width=0.9\linewidth]{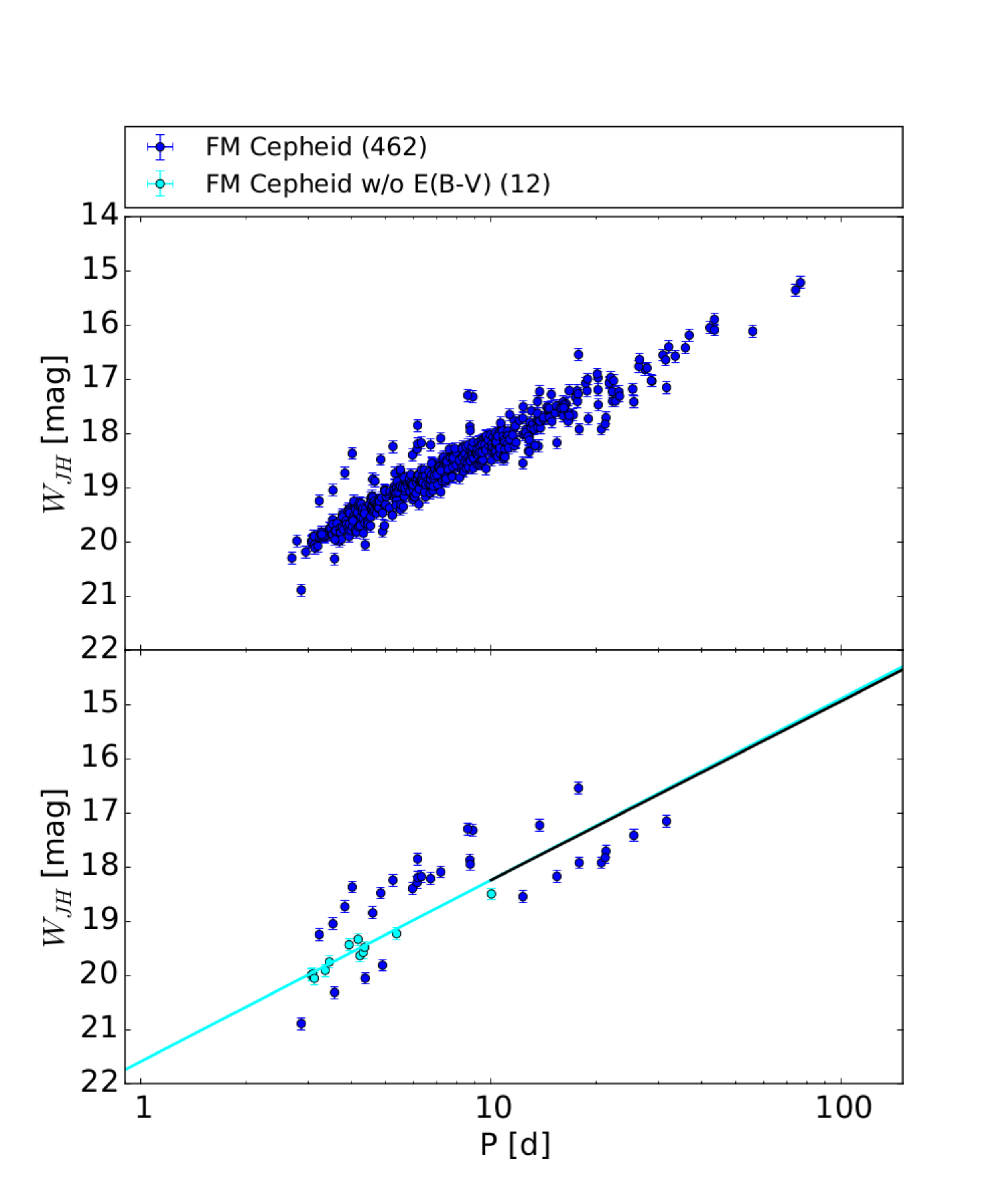}
\caption{Sample III $\mathrm{W}_{\mathrm{JH}}$ PLR \textit{Top panel}: Complete sample before the clipping (462 FM Cepheids) \textit{Bottom panel}: Cepheids that are clipped (32 FM Cepheids) or rejected because there is no color excess (12 FM Cepheids). Note that an error estimate of the mean magnitude correction was added quadratically to the photometric error. Compared to sample I shown in figure \ref{fig_clipping-sample-I-W_JH} the same number of Cepheids is excluded.
\label{fig_clipping-sample-III-W_JH}}
\end{figure*}

\begin{figure*}
\centering
\includegraphics[width=0.9\linewidth]{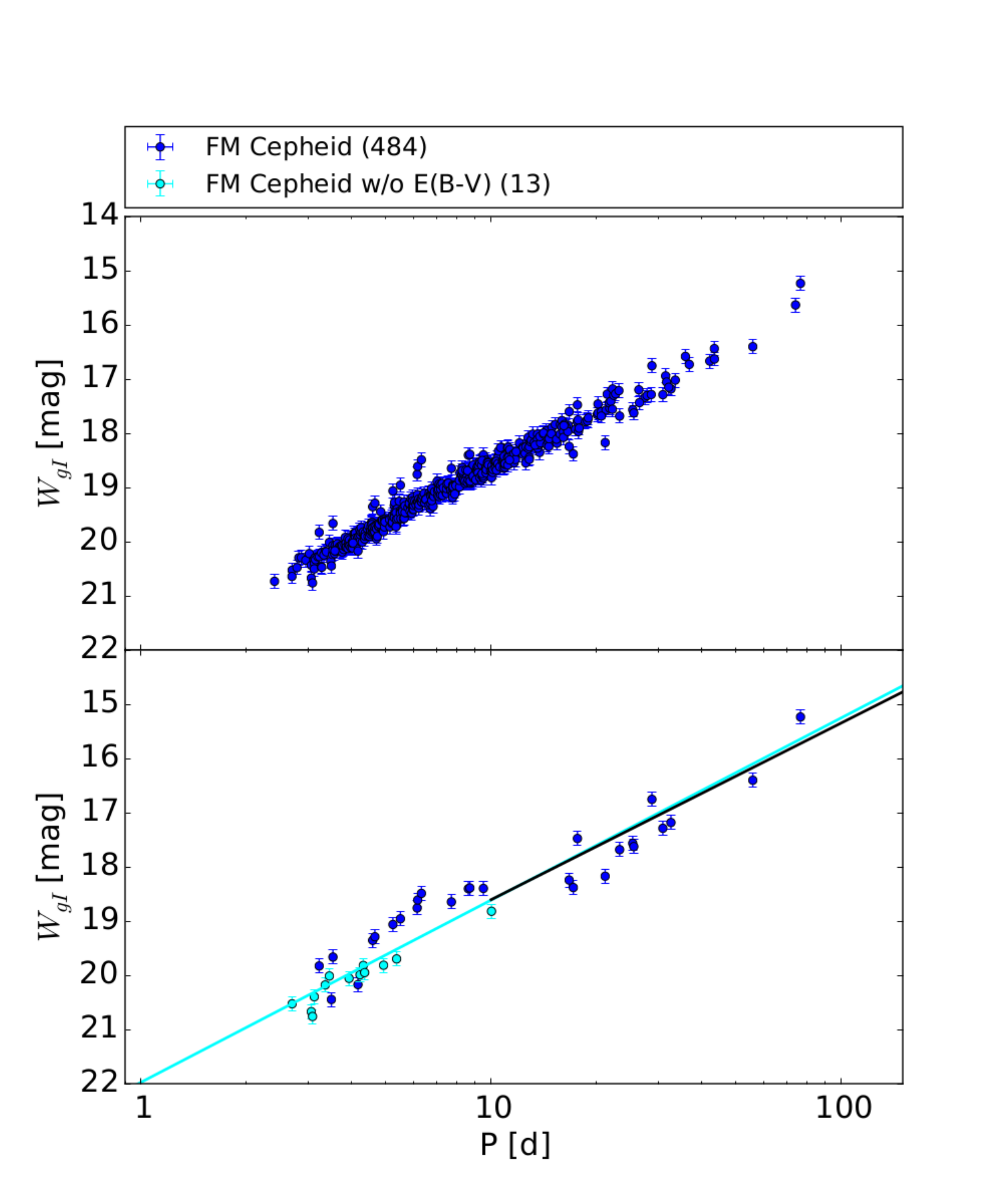}
\caption{Sample III $\mathrm{W}_{\mathrm{gI}}$ PLR \textit{Top panel}: Complete sample before the clipping (484 FM Cepheids) \textit{Bottom panel}: Cepheids that are clipped (27 FM Cepheids) or rejected because there is no color excess (13 FM Cepheids). Note that an error estimate of the mean magnitude correction was added quadratically to the photometric error. Compared to sample I shown in figure \ref{fig_clipping-sample-I-W_gI} more Cepheids are excluded.
\label{fig_clipping-sample-III-W_gI}}
\end{figure*}

Figures \ref{fig_clipping-sample-I-W_JH} and \ref{fig_clipping-sample-I-W_gI} for sample I, figures \ref{fig_clipping-sample-II-W_JH} and \ref{fig_clipping-sample-II-W_gI} for sample II and figures \ref{fig_clipping-sample-III-W_JH} and \ref{fig_clipping-sample-III-W_gI} for sample III, respectively show in the top panels that the PLRs have outliers. As already discussed in K15 the reason for outliers are blending, crowding, extinction, misidentification and misclassification. We will not discuss those here, but refer the interested reader to section 3 in K15, where we discuss the reasons of outliers. We perform outlier clipping using the method we established in K15. The method is an iterative $\kappa$-$\sigma$ clipping with the median absolute deviation of the residuals as the magnitude error. This clipping procedure, that is described in detail in section 3 in K15 is very robust and was later applied by \citet{2016ApJ...826...56R} and \citet{2016ApJ...830...10H}. The outlier rejection is performed for each sample in both of the two Wesenheit magnitudes separately, i.e. in  $\mathrm{W}_{\mathrm{JH}}$ and $\mathrm{W}_{\mathrm{gI}}$ with a $\kappa = 4$. Technically it would be better to require that each Cepheid is not rejected in any of the two Wesenheit Period-Luminosity relations (or alternatively construct a Wesenheit using three different bands). This would of course reduce the sample only to Cepheids that have g, I, J and H band data simultaneously. The reason we treat each filter pair individually is that we have random phased data. This means that the Cepheid will be at a different phase in its light curve in $\mathrm{W}_{\mathrm{JH}}$ than in $\mathrm{W}_{\mathrm{gI}}$ due to the observing strategy. So in order not to introduce any bias and so that our data is more comparable to the literature that typically has data from two bands, we perform the clipping for the ACS/WFC data and WFC3/IR data individually. We also exclude Cepheids that are outside or in a gap of the \citet{2009A&A...507..283M} map, so as to not introduce any systematic effect. But as can be seen in the figures showing the excluded Cepheids, they do still follow the PLRs although the color excess has been set to zero.

The bottom panels of the aforementioned figures show the Cepheids that are not part of the respective sample. These include the Cepheids excluded by the outlier rejection and all the UN Cepheids, as well as the T2 Cepehids which shall not be subject of this paper. For $\mathrm{W}_{\mathrm{JH}}$ in sample I (figure \ref{fig_clipping-sample-I-W_JH}) there are only 32 FM Cepheids and 8 FO Cepheids clipped (less than 10\% of the sample). Most of the outliers below 10 days are FM Cepheids that are on the FO Wesenheit Period-Luminosity relation, which points to misclassification of the type. The type classification is using ground based data and while we improved the classification in K18a, there is a region in the amplitude ratio diagram that is both populated by FM Cepheids and FO Cepheids that makes the classification difficult. Crowding and blending additionally distort the amplitude ratio on which the differentiation of the FM and FO Cepheids rests. In fact most of the UN Cepheids do not have a type assigned because they reside inside this transitional region between FM and FO Cepheids in the amplitude ratio digram. Nevertheless most of the UN Cepheids follow the FM or FO PLR, which means that excluding them due to the Cepheid type uncertainty does not bias the sample. In $\mathrm{W}_{\mathrm{gI}}$ in sample I (figure \ref{fig_clipping-sample-I-W_gI}) 21 FM Cepheids and 4 FO Cepheids that are clipped, which means that this case even less than 5\% of the sample are rejected. Figure \ref{fig_clipping-sample-I-W_JH} shows for $\mathrm{W}_{\mathrm{JH}}$ in sample II the 40 FM Cepheids and 12 FO Cepheids that are clipped. The 43 objects where we could not determine which source the Cepheid is are also shown labeled with bad quality flag (see section \ref{samples}). Figure \ref{fig_clipping-sample-II-W_gI} shows for $\mathrm{W}_{\mathrm{gI}}$ in sample II the 29 FM Cepheids and 7 FO Cepheids that are rejected as well as the 45 objects where the source identification was uncertain. A larger percentage of objects are rejected in sample II compared to sample I, which is not surprising since sample I is more homogeneous than sample II. Nevertheless the outlier rejection works well for both samples.

For sample III figure \ref{fig_clipping-sample-III-W_JH} for $\mathrm{W}_{\mathrm{JH}}$ shows the 32 FM Cepheids that are clipped, while figure \ref{fig_clipping-sample-III-W_gI} for $\mathrm{W}_{\mathrm{gI}}$ shows the 27 clipped FM Cepheids. As described in the previous section an error estimate of the mean magnitude correction was added quadratically to the photometric error of sample III. 

\section{Results \label{results}}

\begin{figure*}
\centering
\includegraphics[width=0.9\linewidth]{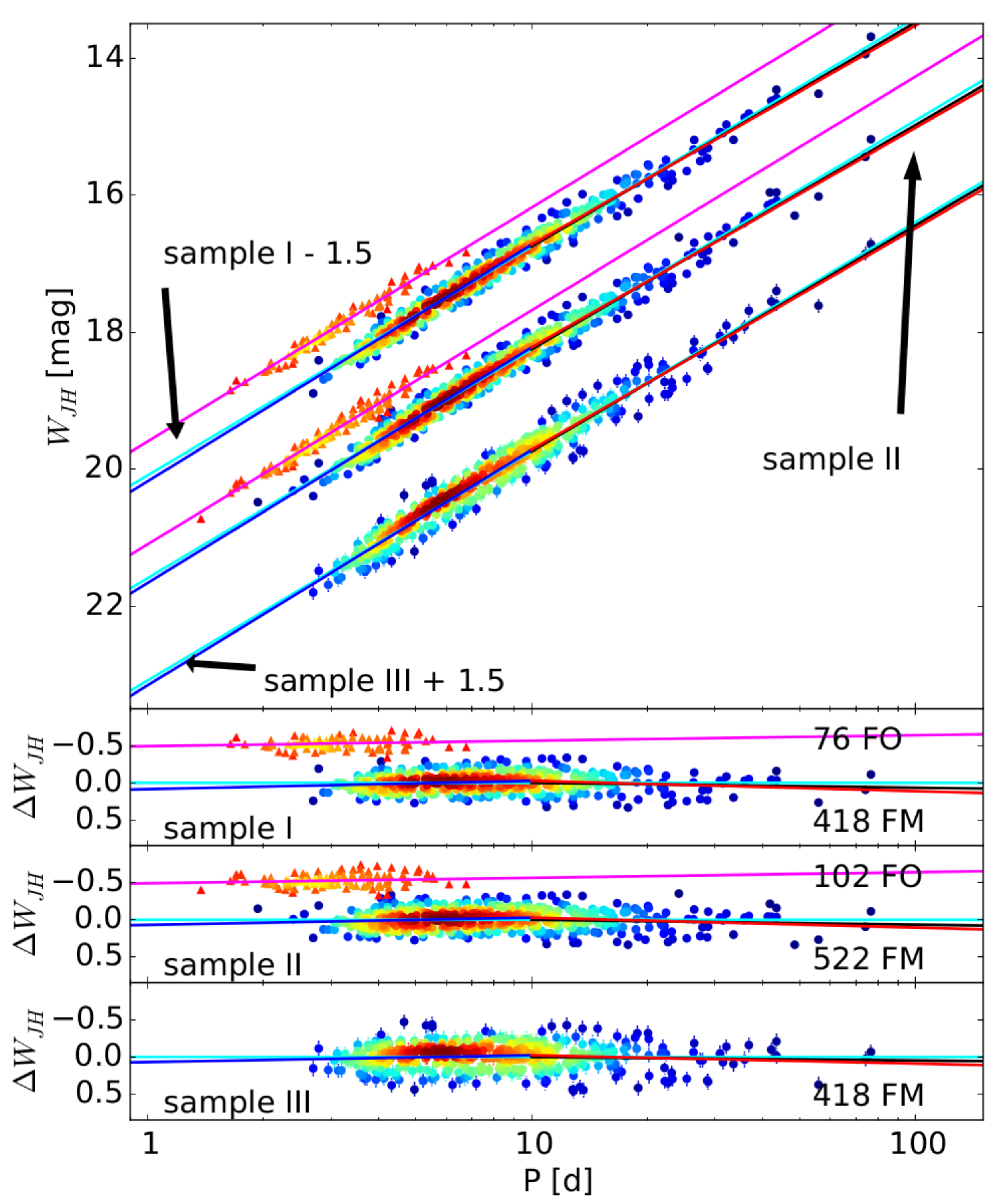}
\caption{\textit{Top panel}: Period-Luminosity relations for the three samples in the $\mathrm{W}_{\mathrm{JH}}$ band. The FM Cepheids are shown as circles, while the FO Cepheids are shown as triangles. The PLRs of the different samples have been offset by 1.5 mag relative to sample II for better visibility. The color is assigned according to the density of Cepheids in this part of the PLR by using the kernel density estimate.
The cyan lines show the best fits to the FM Cepheids (sample I: \#1 in table \ref{table_PLRs_W_sample_I}, sample II: \#1 in table \ref{table_PLRs_W_sample_II}, sample III: \#1 in table \ref{table_PLRs_W_sample_III}), the black lines show the fits to the long period Cepheids (sample I: \#2 in table \ref{table_PLRs_W_sample_I}, sample II: \#2 in table \ref{table_PLRs_W_sample_II}, sample III: \#2 in table \ref{table_PLRs_W_sample_III}) and the magenta lines are the best fits for the FO Cepheids (sample I: \#3 in table \ref{table_PLRs_W_sample_I}, sample II: \#3 in table \ref{table_PLRs_W_sample_II}). The red and blue lines are the broken slope fits (sample I: \#1 in table \ref{table_PLRs_broken_sample_I}, sample II: \#1 in table \ref{table_PLRs_broken_sample_II}, sample III: \#1 in table \ref{table_PLRs_broken_sample_III}). The residuals are shown in the bottom panels.
\label{fig_W_JH}}
\end{figure*}

\begin{figure*}
\centering
\includegraphics[width=0.9\linewidth]{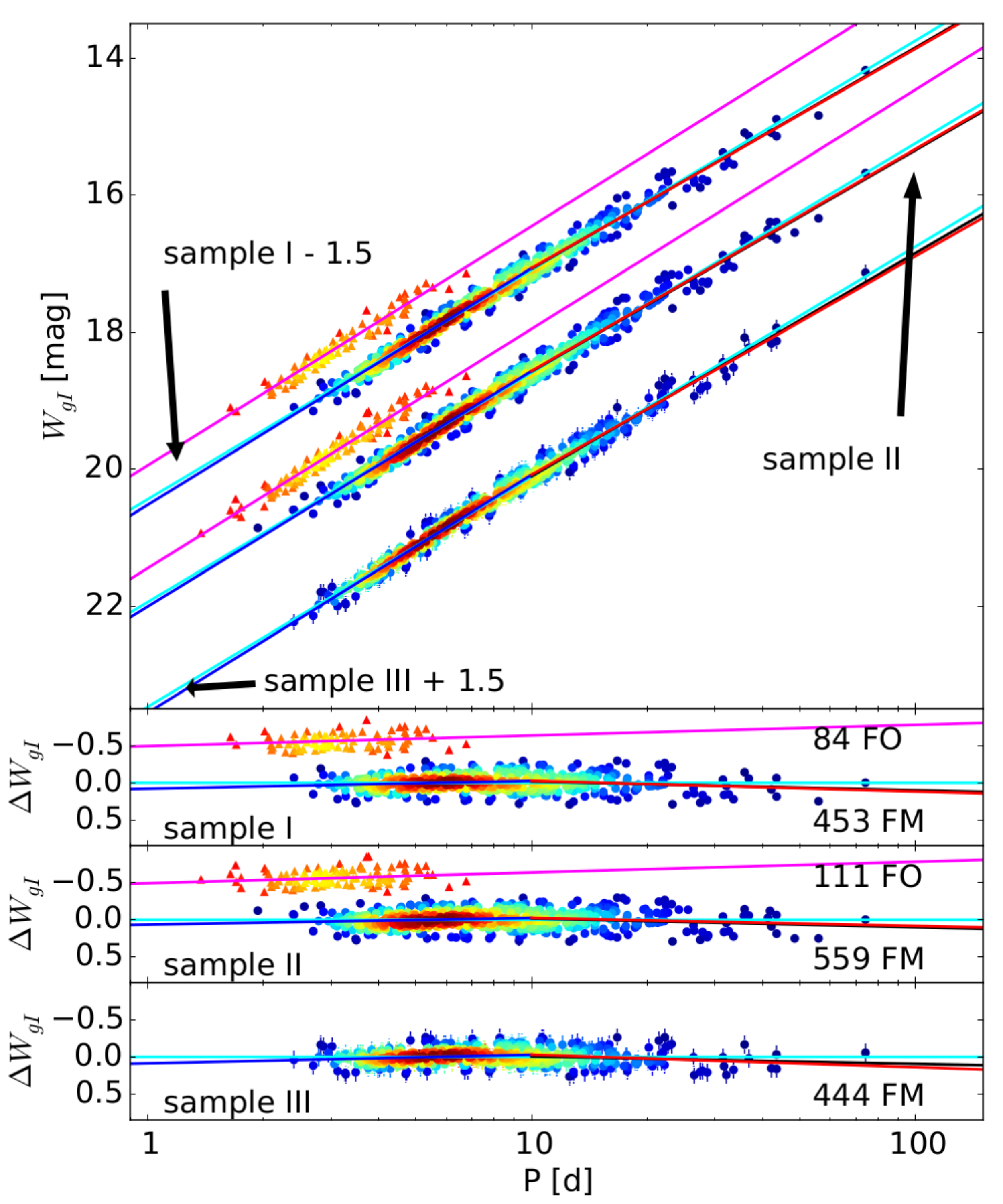}
\caption{\textit{Top panel}: Period-Luminosity relations for the three samples in the $\mathrm{W}_{\mathrm{gI}}$ band. The FM Cepheids are shown as circles, while the FO Cepheids are shown as triangles. The PLRs of the different samples have been offset by 1.5 mag relative to sample II for better visibility. The color is assigned according to the density of Cepheids in this part of the PLR by using the kernel density estimate.
The cyan lines show the best fits to the FM Cepheids (sample I: \#4 in table \ref{table_PLRs_W_sample_I}, sample II: \#4 in table \ref{table_PLRs_W_sample_II}, sample III: \#3 in table \ref{table_PLRs_W_sample_III}), the black lines show the fits to the long period Cepheids (sample I: \#5 in table \ref{table_PLRs_W_sample_I}, sample II: \#5 in table \ref{table_PLRs_W_sample_II}, sample III: \#4 in table \ref{table_PLRs_W_sample_III}) and the magenta lines are the best fits for the FO Cepheids (sample I: \#6 in table \ref{table_PLRs_W_sample_I}, sample II: \#6 in table \ref{table_PLRs_W_sample_II}). The red and blue lines are the broken slope fits (sample I: \#4 in table \ref{table_PLRs_broken_sample_I}, sample II: \#4 in table \ref{table_PLRs_broken_sample_II}, sample III: \#4 in table \ref{table_PLRs_broken_sample_III}). The residuals are shown in the bottom panels.
\label{fig_W_gI}}
\end{figure*}

\begin{figure*}
\centering
\includegraphics[width=0.9\linewidth]{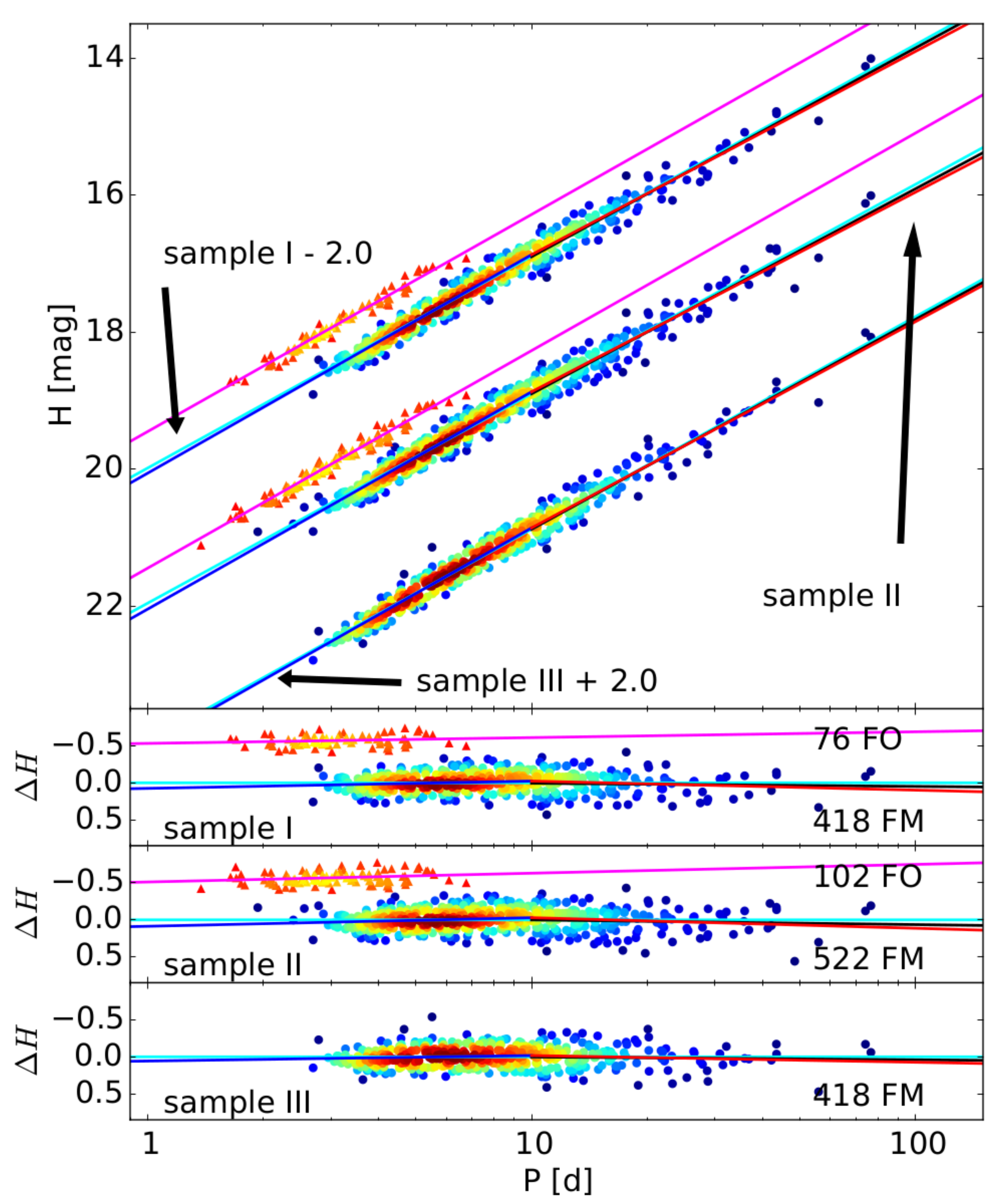}
\caption{\textit{Top panel}: Period-Luminosity relations for the three samples in the H band. The FM Cepheids are shown as circles, while the FO Cepheids are shown as triangles. The PLRs of the different samples have been offset by 2.0 mag relative to sample II for better visibility. The color is assigned according to the density of Cepheids in this part of the PLR by using the kernel density estimate.
The cyan lines show the best fits to the FM Cepheids (sample I: \#1 in table \ref{table_PLRs_sample_I}, sample II: \#1 in table \ref{table_PLRs_sample_II}, sample III: \#1 in table \ref{table_PLRs_sample_III}), the black lines show the fits to the long period Cepheids (sample I: \#2 in table \ref{table_PLRs_sample_I}, sample II: \#2 in table \ref{table_PLRs_sample_II}, sample III: \#2 in table \ref{table_PLRs_sample_III}) and the magenta lines are the best fits for the FO Cepheids (sample I: \#3 in table \ref{table_PLRs_sample_I}, sample II: \#3 in table \ref{table_PLRs_sample_II}). The red and blue lines are the broken slope fits (sample I: \#2 in table \ref{table_PLRs_broken_sample_I}, sample II: \#2 in table \ref{table_PLRs_broken_sample_II}, sample III: \#2 in table \ref{table_PLRs_broken_sample_III}). The residuals are shown in the bottom panels.
\label{fig_H}}
\end{figure*}

\begin{figure*}
\centering
\includegraphics[width=0.9\linewidth]{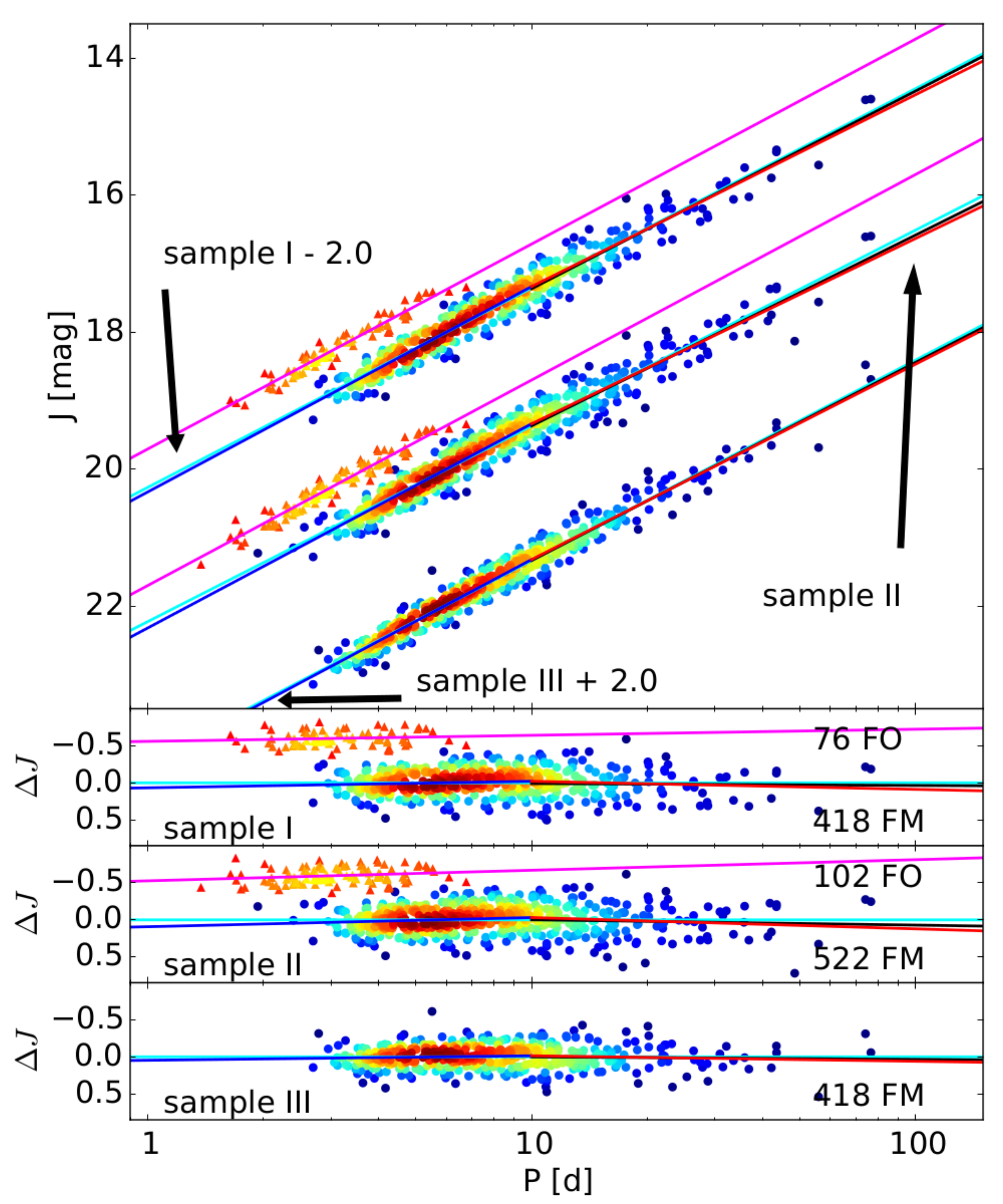}
\caption{\textit{Top panel}: Period-Luminosity relations for the three samples in the J band. The FM Cepheids are shown as circles, while the FO Cepheids are shown as triangles. The PLRs of the different samples have been offset by 2.0 mag relative to sample II for better visibility. The color is assigned according to the density of Cepheids in this part of the PLR by using the kernel density estimate.
The cyan lines show the best fits to the FM Cepheids (sample I: \#4 in table \ref{table_PLRs_sample_I}, sample II: \#4 in table \ref{table_PLRs_sample_II}, sample III: \#3 in table \ref{table_PLRs_sample_III}), the black lines show the fits to the long period Cepheids (sample I: \#5 in table \ref{table_PLRs_sample_I}, sample II: \#5 in table \ref{table_PLRs_sample_II}, sample III: \#4 in table \ref{table_PLRs_sample_III}) and the magenta lines are the best fits for the FO Cepheids (sample I: \#6 in table \ref{table_PLRs_sample_I}, sample II: \#6 in table \ref{table_PLRs_sample_II}). The red and blue lines are the broken slope fits (sample I: \#3 in table \ref{table_PLRs_broken_sample_I}, sample II: \#3 in table \ref{table_PLRs_broken_sample_II}, sample III: \#3 in table \ref{table_PLRs_broken_sample_III}). The residuals are shown in the bottom panels.
\label{fig_J}}
\end{figure*}

\begin{figure*}
\centering
\includegraphics[width=0.9\linewidth]{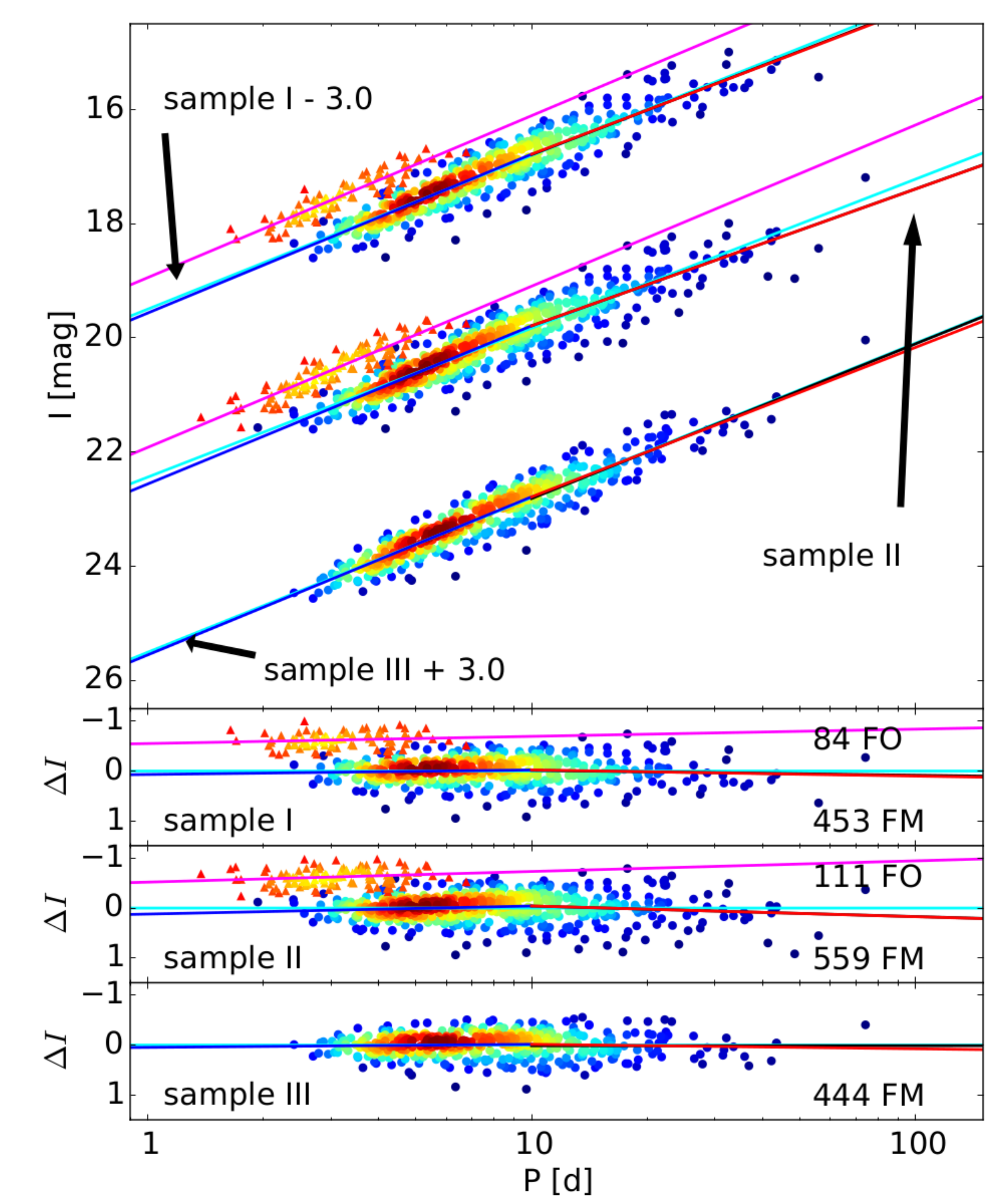}
\caption{\textit{Top panel}: Period-Luminosity relations for the three samples in the I band. The FM Cepheids are shown as circles, while the FO Cepheids are shown as triangles. The PLRs of the different samples have been offset by 3.0 mag relative to sample II for better visibility. The color is assigned according to the density of Cepheids in this part of the PLR by using the kernel density estimate.
The cyan lines show the best fits to the FM Cepheids (sample I: \#7 in table \ref{table_PLRs_sample_I}, sample II: \#7 in table \ref{table_PLRs_sample_II}, sample III: \#5 in table \ref{table_PLRs_sample_III}), the black lines show the fits to the long period Cepheids (sample I: \#8 in table \ref{table_PLRs_sample_I}, sample II: \#8 in table \ref{table_PLRs_sample_II}, sample III: \#6 in table \ref{table_PLRs_sample_III}) and the magenta lines are the best fits for the FO Cepheids (sample I: \#9 in table \ref{table_PLRs_sample_I}, sample II: \#9 in table \ref{table_PLRs_sample_II}). The red and blue lines are the broken slope fits (sample I: \#5 in table \ref{table_PLRs_broken_sample_I}, sample II: \#5 in table \ref{table_PLRs_broken_sample_II}, sample III: \#5 in table \ref{table_PLRs_broken_sample_III}). The residuals are shown in the bottom panels.
\label{fig_I}}
\end{figure*}

\begin{figure*}
\centering
\includegraphics[width=0.9\linewidth]{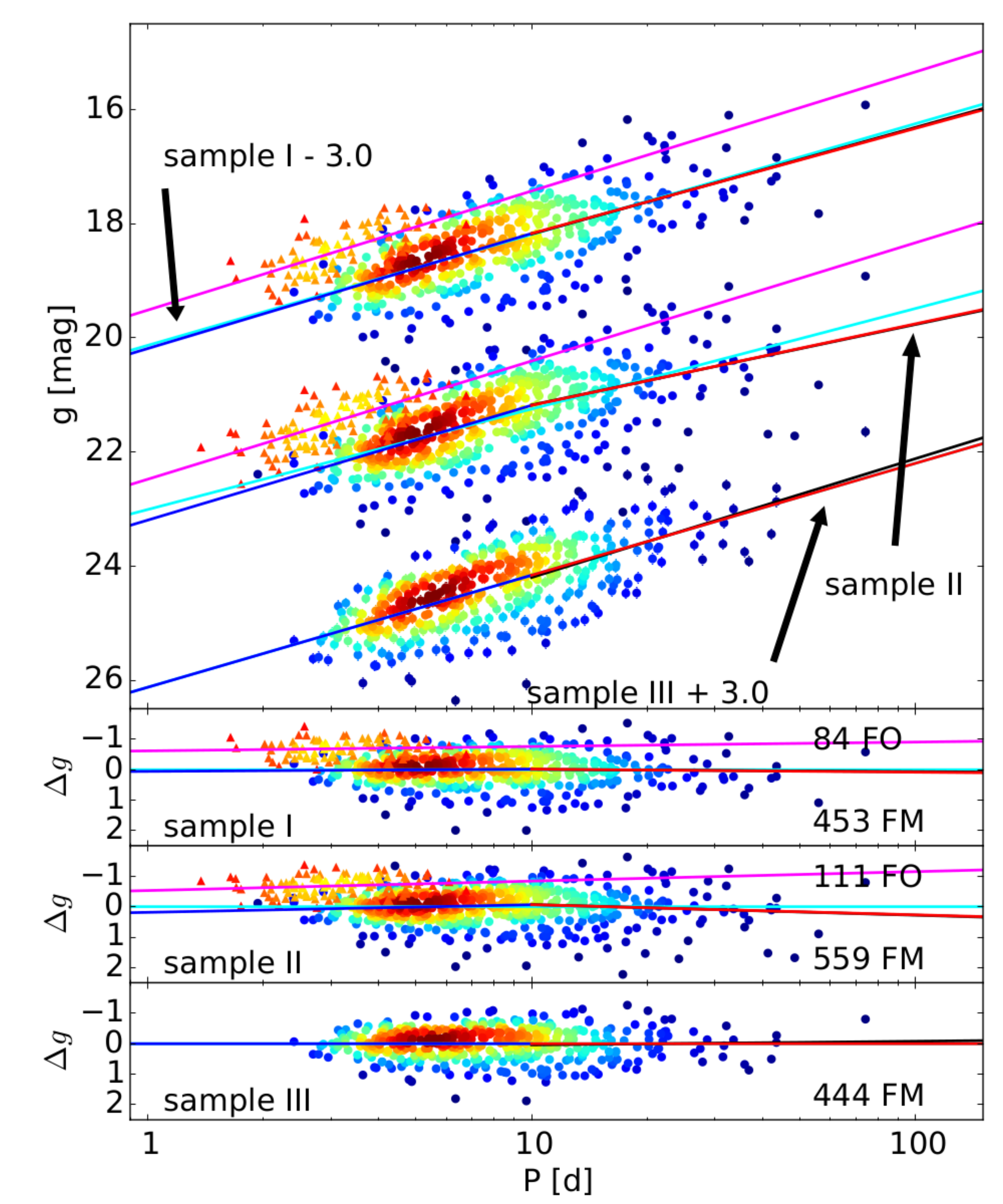}
\caption{\textit{Top panel}: Period-Luminosity relations for the three samples in the g band. The FM Cepheids are shown as circles, while the FO Cepheids are shown as triangles. The PLRs of the different samples have been offset by 3.0 mag relative to sample II for better visibility. The color is assigned according to the density of Cepheids in this part of the PLR by using the kernel density estimate.
The cyan lines show the best fits to the FM Cepheids (sample I: \#10 in table \ref{table_PLRs_sample_I}, sample II: \#10 in table \ref{table_PLRs_sample_II}, sample III: \#7 in table \ref{table_PLRs_sample_III}), the black lines show the fits to the long period Cepheids (sample I: \#11 in table \ref{table_PLRs_sample_I}, sample II: \#11 in table \ref{table_PLRs_sample_II}, sample III: \#8 in table \ref{table_PLRs_sample_III}) and the magenta lines are the best fits for the FO Cepheids (sample I: \#12 in table \ref{table_PLRs_sample_I}, sample II: \#12 in table \ref{table_PLRs_sample_II}). The red and blue lines are the broken slope fits (sample I: \#6 in table \ref{table_PLRs_broken_sample_I}, sample II: \#6 in table \ref{table_PLRs_broken_sample_II}, sample III: \#6 in table \ref{table_PLRs_broken_sample_III}). The residuals are shown in the bottom panels.
\label{fig_g}}
\end{figure*}

The Period-Luminosity relations for the three different samples are shown in figures \ref{fig_W_JH}, \ref{fig_W_gI}, \ref{fig_H}, \ref{fig_J}, \ref{fig_I} and \ref{fig_g}. The corresponding fit parameters are found in tables \ref{table_PLRs_W_sample_I}, \ref{table_PLRs_sample_I}, \ref{table_PLRs_broken_sample_I}, \ref{table_PLRs_W_sample_II}, \ref{table_PLRs_sample_II}, \ref{table_PLRs_broken_sample_II}, \ref{table_PLRs_W_sample_III}, \ref{table_PLRs_sample_III} and \ref{table_PLRs_broken_sample_III}. Surprisingly, the dispersion in the $\mathrm{W}_{\mathrm{JH}}$, J and H PLRs is smaller than in K15, although the sample is larger. The dispersion in $\mathrm{W}_{\mathrm{gI}}$ is smaller than in $\mathrm{W}_{\mathrm{JH}}$ although the dispersions in the I and g band are much larger than in the J and H bands. Also, as expected \citep{2012ApJ...758...24F} the dispersion in general is larger for shorter wavelengths. The dispersion in sample II is larger than in sample I, but sample II contains more Cepheids. The obtained slopes in sample II are, consistently over all wavelengths, slightly steeper compared to sample I. Overall the dispersions improve by applying the mean magnitude correction with the exception of the $\mathrm{W}_{\mathrm{JH}}$ dispersion and the dispersion in the H band PLR. The mean magnitude correction in the H band does not work in contrast to that in the J band and therefore the Wesenheit is affected strongly. Since the method of correction is the same in the H and J band the most likely reason for the problem in the H band is that the phase lag determined by \citet{2015A&A...576A..30I} in the H band is not compatible with that determined in the J band. The amplitudes in the g and I band are larger than, e.g., in the J band (see e.g. \citet{1991PASP..103..933M}) and therefore the correction should have a larger impact in those bands, which is not the case. The multiple epochs in the g and I band seem to provide a good estimate of the mean magnitude even without the correction. The slopes between sample III and sample I agree within the errors although the slopes from sample III are consistently shallower than in sample I. In the J and H band the $\log(P) > 1$ subsamples again have a steeper slope than the complete sample as predicted by \citet{2010ApJ...715..277B}. In the I band the slopes are almost the same and in the g band the complete sample has a steeper slope when comparing sample III with sample I.

\begin{figure*}
\centering
\includegraphics[width=0.85\linewidth]{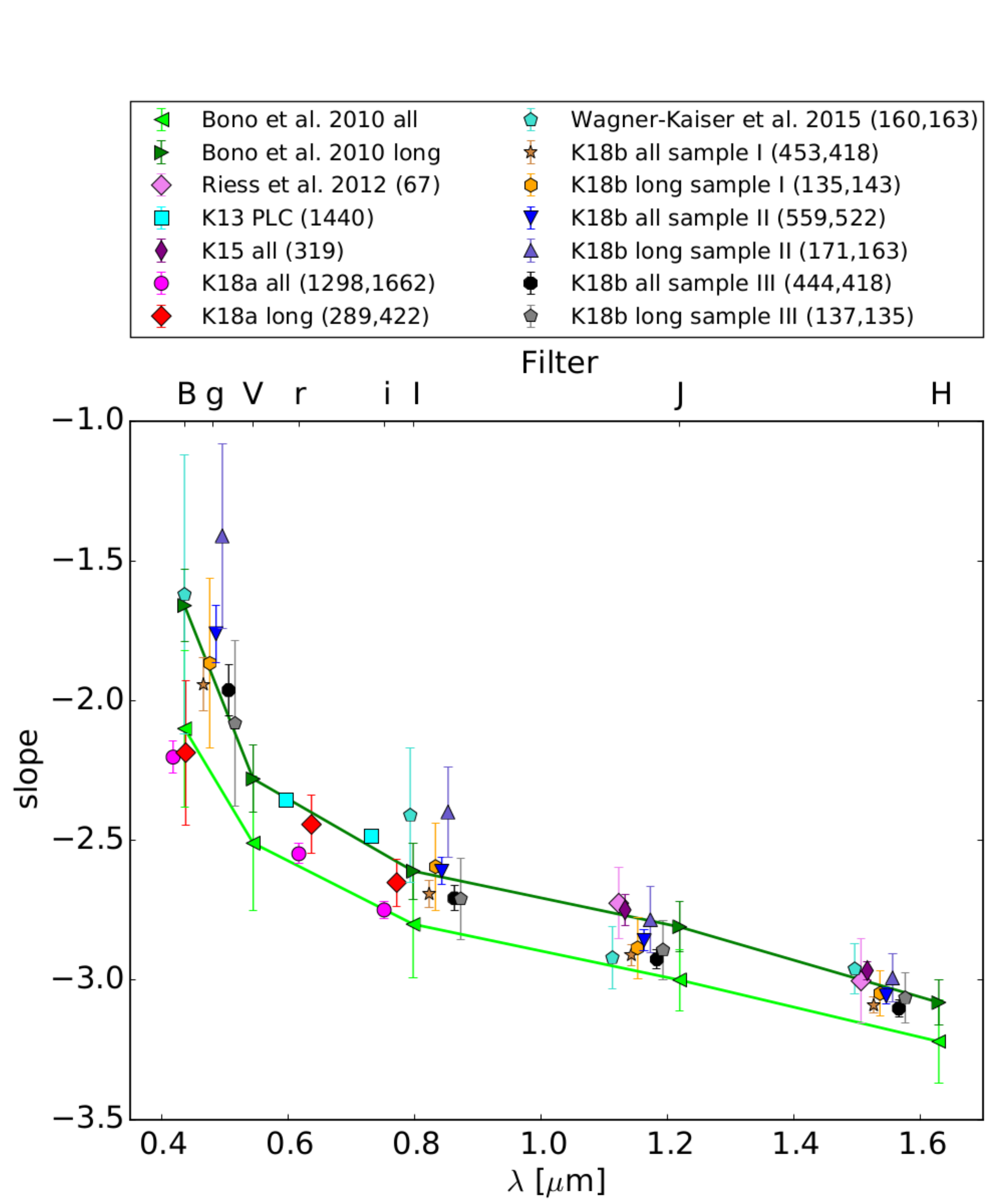}
\caption{PLR slope comparison for M31 with the literature for different wavelengths. Shown are the theoretical \citet{2010ApJ...715..277B} predictions, the slopes obtained by \citet{2015MNRAS.451..724W} and \citet{2012ApJ...745..156R}. Also shown are the slopes from \citet{K13}, \citet{K15} and \citet{K18a}. The fits to the FM Cepheids shown in table \ref{table_PLRs_sample_I}, table \ref{table_PLRs_sample_II} and table \ref{table_PLRs_sample_III} are also shown. We find that as predicted by \citet{2010ApJ...715..277B} our long period slopes are steeper than the slopes for the complete sample, although all those slopes are steeper than the prediction. We do not observe that the slope is almost the same in all the long wavelengths seen by \citet{2015MNRAS.451..724W}. Our slopes here are also steeper than the slopes we obtained from ground-based observations (K13 and K18a), but shallower than the slopes obtained by R12 and our previous work in K15. In round brackets in the legend the sample size of the different samples is given. In the cases where two numbers are given, the sample sizes are different for different wavelengths. 
\label{fig_slopelambda_M31}}
\end{figure*}

\begin{figure*}
\centering
\includegraphics[width=0.95\linewidth]{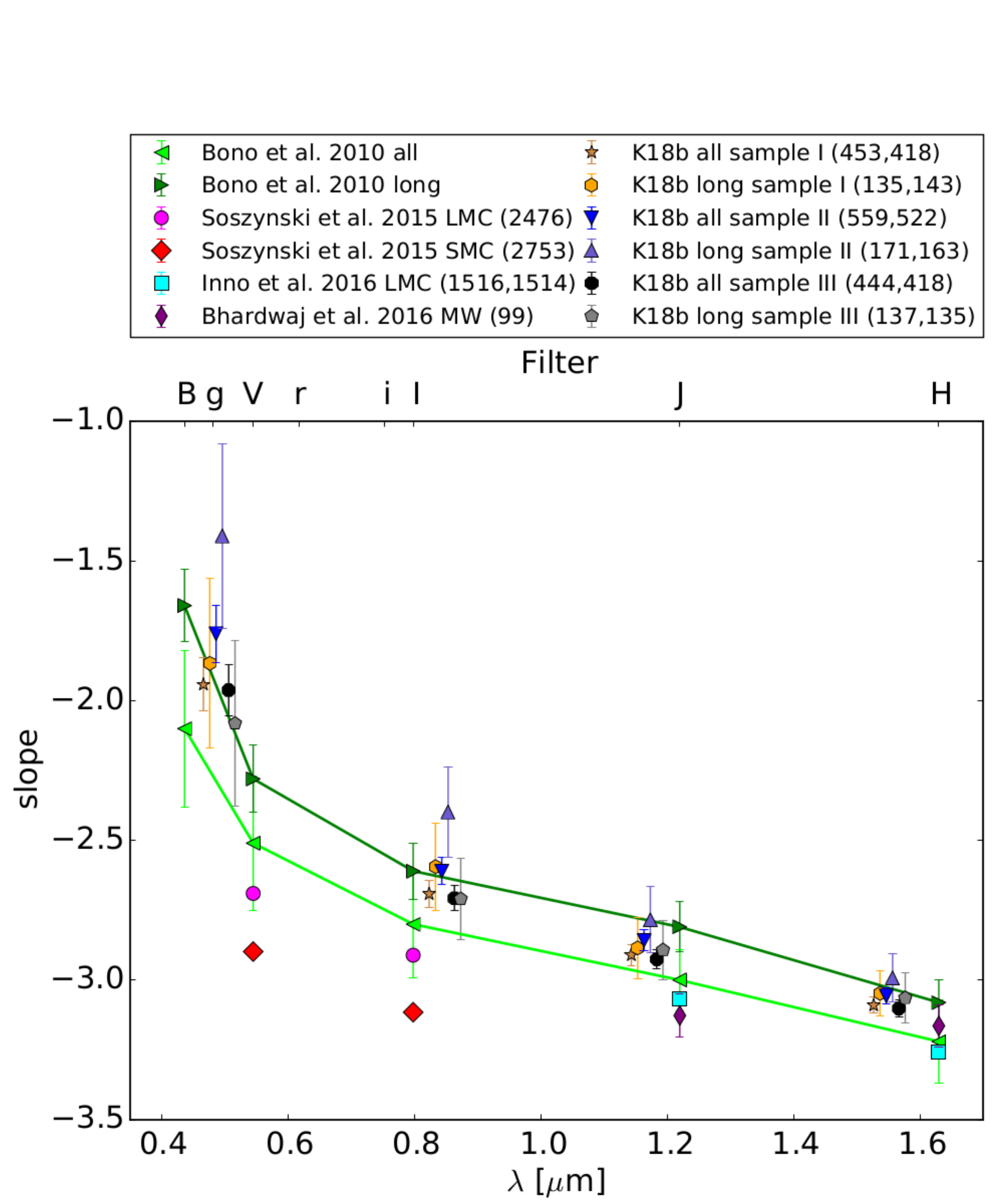}
\caption{PLR slope comparison between our M31 slopes (see also figure \ref{fig_slopelambda_M31}) and literature slopes of other galaxies. The slopes for LMC and SMC in \citet{2015AcA....65..297S} as well as the LMC slope in \citet{2016ApJ...832..176I} and the slope for the Milky Way (MW) in \citet{2016AJ....151...88B} are all consistently shallower than our values obtained for M31. The difference in the slope compared to our sloped gets smaller the larger the metallicity gets, i.e. the slopes have a larger difference for the SMC than for the MW. This points to a metallicity dependence of the slope. 
\label{fig_slopelambda}}
\end{figure*}

A comparison of the slopes with literature values is shown in figures \ref{fig_slopelambda_M31} and \ref{fig_slopelambda}. The slopes obtained here are steeper than the \citet{2010ApJ...715..277B} predictions for the complete period range, but consistent with the prediction that the $\log(P) > 1$ subsamples have a steeper slope than the complete samples. Our slopes here are also steeper than the slopes we obtained from ground-based observations (K13 and K18a), but shallower than the slopes obtained by R12 and our previous work in K15. We do not observe that the slope is almost the same in all the long wavelengths seen by \citet{2015MNRAS.451..724W}. The slopes for LMC and SMC in \citet{2015AcA....65..297S} as well as the LMC slope in \citet{2016ApJ...832..176I} and the slope for the Milky way (MW) in \citet{2016AJ....151...88B} are all consistently shallower than our values obtained for M31. The difference in the slope compared to our sloped gets smaller the larger the metallicity gets, i.e. the slopes have a larger difference for the SMC than for the MW. This points to a metallicity dependence of the slope.

\begin{figure*}
\centering
\includegraphics[width=\linewidth]{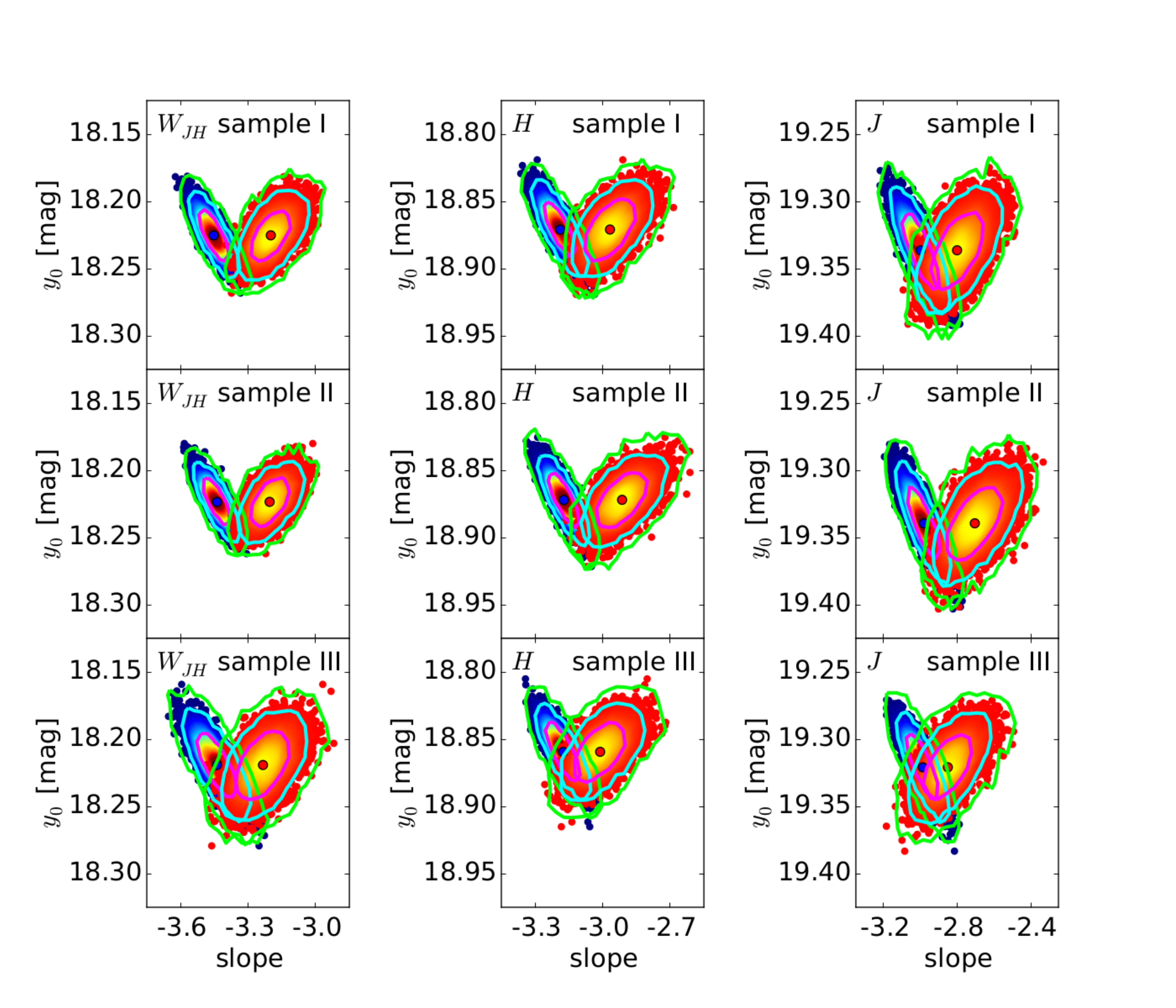}
\caption{Bootstrapping of the broken slope fit for the $\mathrm{W}_{\mathrm{JH}}$, J and H PLRs. Shown is the common suspension point $y_0$ vs. the slope. The points are colored according to the kernel density estimate. The $1\sigma$, $2\sigma$ and $3\sigma$ contour lines are shown as solid lines. The overlapping contour lines show that there is no evidence for a broken slope.
\label{fig_boot_NIR}}
\end{figure*}

\begin{figure*}
\centering
\includegraphics[width=\linewidth]{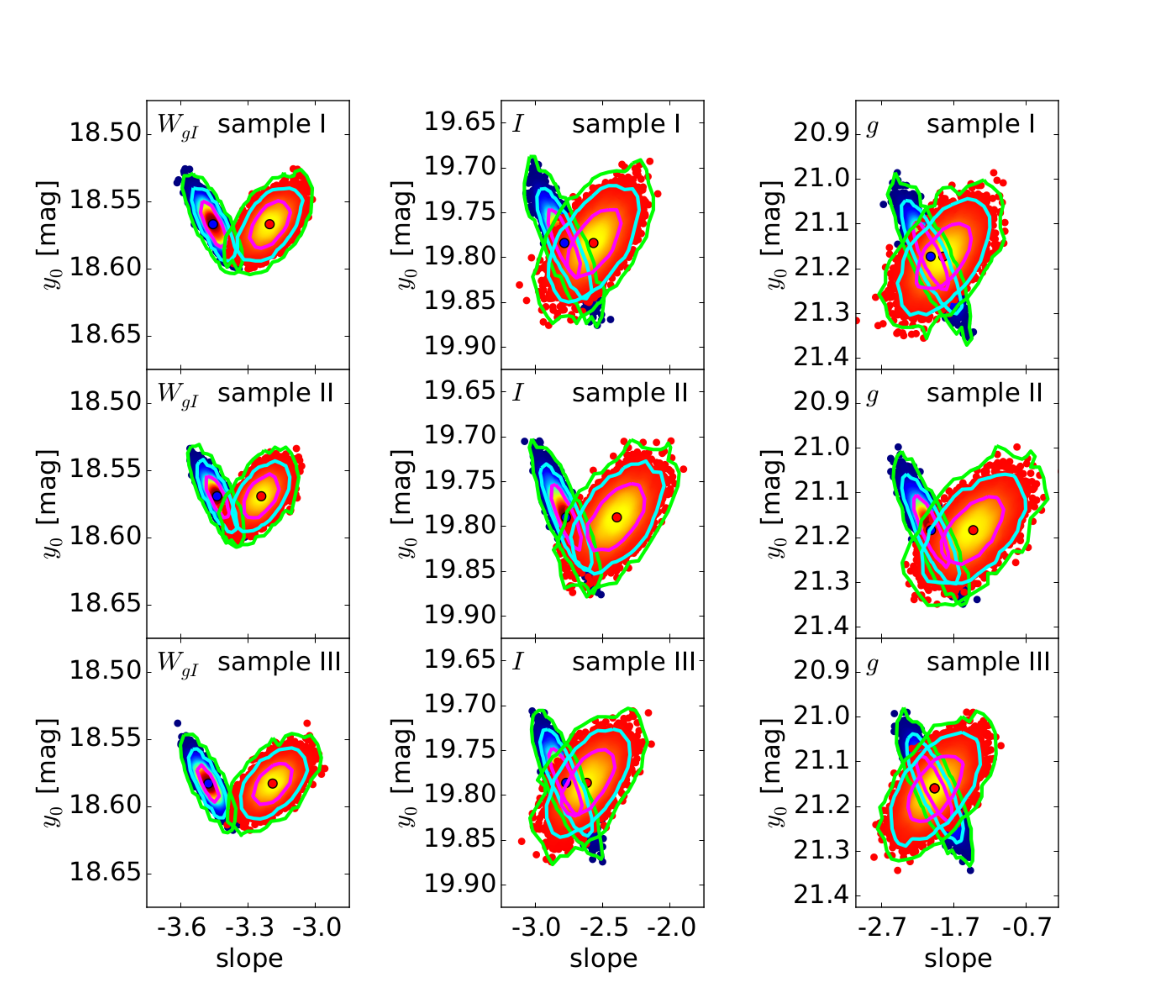}
\caption{Bootstrapping of the broken slope fit for the $\mathrm{W}_{\mathrm{gI}}$, g and I PLRs. Shown is the common suspension point $y_0$ vs. the slope. The points are colored according to the kernel density estimate. The $1\sigma$, $2\sigma$ and $3\sigma$ contour lines are shown as solid lines. The overlapping contour lines show that there is no evidence for a broken slope.
\label{fig_boot_Opt}}
\end{figure*}

In order to investigate, whether there is a broken slope, we fit two slopes $b_{log(P)\leq1}$ and $b_{\log(P)>1}$ and a common suspension point $a_{\log(P)=1}$ (also called $y_0$). We perform 10000 bootstrapping runs in order to see if the two slopes are significantly different. The bootstrapping also helps to negate the effect a few points at very long or short periods could have on the fit. The bootstrapped broken slope fits show in figures \ref{fig_boot_NIR} and \ref{fig_boot_Opt} that the slopes get shallower for longer wavelengths, while the dispersion gets smaller. The main result is however that the contour lines overlap and therefore we do not see a clear evidence for a broken slope. The least overlap is seen in the Wesenheit PLRs, but in the other bands we see a clear overlap contradictory to what we found in K15. This is rather surprising since the K15 sample is based on K13 and this sample is included in sample II (also a large fraction of K15 is included in sample I). From the 335 Cepheids in K15 (319 FM and 16 FO Cepheids) 22 are now classified as UN Cepheids and 4 are clipped. The remaining 309 are part of sample II. In fact 221 FM Cepheids and 9 FO Cepheids are also part of sample I. The remaining 73 FM Cepheids and 6 FO Cepheids are not part of sample I and only present in sample II. We test if we can reproduce the result of the broken slope with only the Cepheids that are in K15. We perform this test in the H band since there the broken slope was most significant. We find that the K15 Cepheids that are in sample II show a broken slope (only the 3 $\sigma$ contour lines overlap slightly) and that the K15 Cepheids that are in sample I show no broken slope (contour lines overlap like in figure \ref{fig_boot_NIR}). This very surprising result means that the 73 FM Cepheids that are only in sample II produce the broken slope. In fact the residuals of those 73 FM Cepheids show that those points alone would create an even more significant broken slope. In a next step we investigate why those 73 Cepheids are not part of the K18a sample and therefore part of sample I. We find that for 5 we do not have data in K18a because of stricter masking, 21 have not periods in the \rps and \ips band that are similar on a one percent level and one is cut because of an amplitude criterion (see K18a for more details). The remaining 46 are cut due to the color cut criterion (see K18a for more details on the color cut). About a third are too blue while the rest is too red. Also the color excess of the 46 is larger than the color excess of most of the sample I Cepheids (consistent with the fact that the 46 are inside the spiral arms). The broken slope is also present if we perform no extinction correction. The light curves also do look like typical Cepheid light curves (in fact 38 of the 46 were manually classified as Cepheids). We do not find a reason why these 46 Cepheids are different from the other Cepheids other than their color. But they populate a PLR with a distinct broken slope. Those 46 influence the slopes so much that if combined with other 248 FM Cepheids from K15 that are in sample II, the broken slope is still significant. This is not the case with the complete sample II since the 476 FM Cepheids obscure the effect of the 46 FM Cepheids. The (rather unsatisfactory) reason why the broken slope is sometimes present in the literature and sometimes not seems to be based on a selection effect, where a significant enough presence of this special type of Cepheid in the sample is the cause for the broken slope.  

\begin{figure*}
\centering
\includegraphics[width=0.9\linewidth]{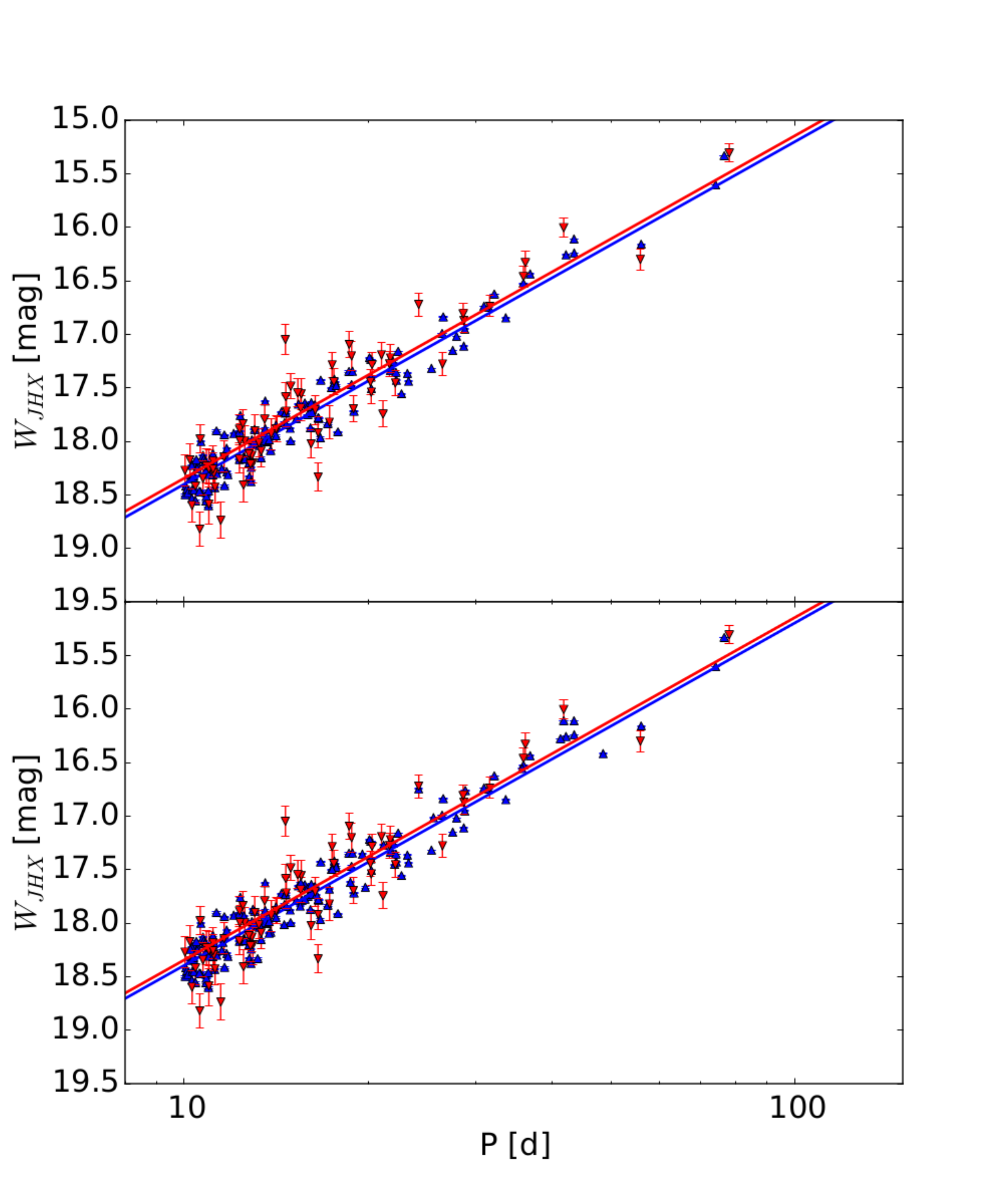}
\caption{Comparison of the R12 sample with sample I (top panel) and sample II (bottom panel). $\mathrm{W}_{\mathrm{JHX}}$ is the color corrected Wesenheit as used in figure 27 in K15. A fixed slope of -3.20 was fitted to both samples so that we can compare with fit \#10 table 3 in R12. 	
For sample I we find an offset of 0.056 mag, corresponding to a 2.6\% larger $\mathrm{H}_0$, while for sample II the offset is 0.050 mag, which results in a 2.4\% larger $\mathrm{H}_0$. If we do not use the photometric errors given in R12, i.e. we do not use them as weight in the fit, the offset decreases to 0.030 mag for sample I (a 1.4\% larger $\mathrm{H}_0$) and 0.025 mag for sample II (a 1.1\% larger $\mathrm{H}_0$). So we observe a sample selection bias on $\mathrm{H}_0$ in the order of the error budget given in \citet{2016ApJ...826...56R}.
\label{fig_H0}}
\end{figure*}

In K15 we found a sample selection bias that changes the value of the Hubble constant by 3.2\%. We check sample I and sample II for this sample selection bias. For the comparison we use fit \#10 in table 3 in R12, where M31 was used as an anchor galaxy to determine the Hubble constant. In order to be able to make the comparison between our samples and the R12 sample, from which we can calculate the impact on $\mathrm{H}_0$, we have to make a transformation to the Wesenheit used in R12 and fit the offset with the slope that was used in R12. The complete procedure is described in detail in section 5 in K15. The comparison is shown in figure \ref{fig_H0}. For sample I we find an offset of 0.056 mag, while for sample II the offset is 0.050 mag. This corresponds to a 2.6\% larger $\mathrm{H}_0$ for sample I and a 2.4\% larger $\mathrm{H}_0$ for sample II. If we do not use the R12 photometric errors in the fit, i.e. assign the same weight to all R12 points, the offset decreases to 0.030 mag for sample I and to 0.025 mag in sample II, i.e. a 1.4\% or 1.1\% larger $\mathrm{H}_0$ respectively. This means that we still find the sample selection bias but not as strong as in K15, although it is still at a level of the error budget given in \citet{2016ApJ...826...56R}. This sample selection estimate covers only the simple case where the PLR slope between the anchor and the other galaxies does not change. But as discussed above the broken slope is caused by a selection effect, so the slope would change from galaxy to galaxy.

\begin{deluxetable}{cccccccc}
\tabletypesize{\scriptsize}
\tablecaption{sample I Wesenheit PLR fit parameters\label{table_PLRs_W_sample_I}}
\tablewidth{0pt}
\tablehead{
\colhead{$\#$} & \colhead{band} & \colhead{type} & \colhead{range} & \colhead{$N_{fit}$} & \colhead{a (log P = 1)} & \colhead{slope b} & \colhead{$\sigma$}
}
\startdata

1 & $\mathrm{W}_{\mathrm{JH}}$ & FM & all &  418 & 18.253 (0.007) & -3.340 (0.025) & 0.131 \\
2 & $\mathrm{W}_{\mathrm{JH}}$ & FM & log(P) $>$ 1 &  135 & 18.252 (0.018) & -3.273 (0.067) & 0.147 \\
3 & $\mathrm{W}_{\mathrm{JH}}$ & FO & all &   76 & 17.684 (0.036) & -3.414 (0.065) & 0.078 \\
4 & $\mathrm{W}_{\mathrm{gI}}$ & FM & all &  453 & 18.594 (0.007) & -3.349 (0.024) & 0.116 \\
5 & $\mathrm{W}_{\mathrm{gI}}$ & FM & log(P) $>$ 1 &  143 & 18.576 (0.015) & -3.230 (0.064) & 0.129 \\
6 & $\mathrm{W}_{\mathrm{gI}}$ & FO & all &   84 & 17.954 (0.045) & -3.493 (0.084) & 0.096 \\
\enddata
\tablecomments{The magnitude errors were set to the same value.
  The errors of the fitted parameters were determined with the bootstrapping method.
}
\end{deluxetable}

\begin{deluxetable}{cccccccc}
\tabletypesize{\scriptsize}
\tablecaption{sample I PLR fit parameters\label{table_PLRs_sample_I}}
\tablewidth{0pt}
\tablehead{
\colhead{$\#$} & \colhead{band} & \colhead{type} & \colhead{range} & \colhead{$N_{fit}$} & \colhead{a (log P = 1)} & \colhead{slope b} & \colhead{$\sigma$}
}
\startdata
1 & H & FM & all & 418 & 18.895 (0.008) & -3.090 (0.028) & 0.138 \\
2 & H & FM & log(P) $>$ 1 & 135 & 18.900 (0.022) & -3.047 (0.084) & 0.164 \\
3 & H & FO & all &  76 & 18.284 (0.038) & -3.169 (0.069) & 0.075 \\
4 & J & FM & all & 418 & 19.358 (0.010) & -2.910 (0.037) & 0.184 \\
5 & J & FM & log(P) $>$ 1 & 135 & 19.367 (0.029) & -2.885 (0.110) & 0.219 \\
6 & J & FO & all &  76 & 18.716 (0.046) & -2.992 (0.088) & 0.100 \\
7 & I & FM & all & 453 & 19.807 (0.014) & -2.692 (0.049) & 0.256 \\
8 & I & FM & log(P) $>$ 1 & 143 & 19.793 (0.040) & -2.594 (0.157) & 0.300 \\
9 & I & FO & all &  84 & 19.110 (0.067) & -2.836 (0.130) & 0.150 \\
10 & g & FM & all & 453 & 21.192 (0.029) & -1.942 (0.096) & 0.519 \\
11 & g & FM & log(P) $>$ 1 & 143 & 21.182 (0.078) & -1.867 (0.303) & 0.602 \\
12 & g & FO & all &  84 & 20.429 (0.121) & -2.087 (0.240) & 0.300 \\
\enddata
\tablecomments{The magnitude errors were set to the same value.
  The errors of the fitted parameters were determined with the bootstrapping method.
}
\end{deluxetable}

\begin{deluxetable}{ccccccc}
\tabletypesize{\scriptsize}
\tablecaption{sample I broken slope PLR fit parameters\label{table_PLRs_broken_sample_I}}
\tablewidth{0pt}
\tablehead{
\colhead{$\#$} & \colhead{band} & \colhead{$N_{fit}$} & \colhead{$b_{\log(P)\leq1}$} & \colhead{$b_{\log(P)>1}$} &\colhead{$a_{\log(P)=1}$} & \colhead{$\sigma$}
}
\startdata
1 & $\mathrm{W}_{\mathrm{JH}}$ &  418 & -3.453 (0.043) & -3.199 (0.057) & 18.225 (0.012) & 0.129 \\
2 & H &  418 & -3.189 (0.047) & -2.967 (0.068) & 18.871 (0.013) & 0.137 \\
3 & J &  418 & -2.998 (0.062) & -2.800 (0.088) & 19.336 (0.017) & 0.183 \\
4 & $\mathrm{W}_{\mathrm{gI}}$ &  453 & -3.456 (0.036) & -3.204 (0.054) & 18.567 (0.010) & 0.114 \\
5 & I &  453 & -2.783 (0.078) & -2.568 (0.122) & 19.784 (0.024) & 0.255 \\
6 & g &  453 & -2.016 (0.157) & -1.841 (0.240) & 21.173 (0.047) & 0.519 \\
\enddata
\tablecomments{The magnitude errors were set to the same value. The errors of the fitted parameters were determined with the bootstrapping method.
}
\end{deluxetable}

\begin{deluxetable}{cccccccc}
\tabletypesize{\scriptsize}
\tablecaption{sample II Wesenheit PLR fit parameters\label{table_PLRs_W_sample_II}}
\tablewidth{0pt}
\tablehead{
\colhead{$\#$} & \colhead{band} & \colhead{type} & \colhead{range} & \colhead{$N_{fit}$} & \colhead{a (log P = 1)} & \colhead{slope b} & \colhead{$\sigma$}
}
\startdata
1 & $\mathrm{W}_{\mathrm{JH}}$ & FM & all &  522 & 18.250 (0.007) & -3.334 (0.024) & 0.134 \\
2 & $\mathrm{W}_{\mathrm{JH}}$ & FM & log(P) $>$ 1 &  163 & 18.252 (0.017) & -3.281 (0.066) & 0.149 \\
3 & $\mathrm{W}_{\mathrm{JH}}$ & FO & all &  102 & 17.674 (0.032) & -3.419 (0.056) & 0.084 \\
4 & $\mathrm{W}_{\mathrm{gI}}$ & FM & all &  559 & 18.591 (0.006) & -3.354 (0.021) & 0.117 \\
5 & $\mathrm{W}_{\mathrm{gI}}$ & FM & log(P) $>$ 1 &  171 & 18.577 (0.014) & -3.264 (0.058) & 0.132 \\
6 & $\mathrm{W}_{\mathrm{gI}}$ & FO & all &  111 & 17.950 (0.039) & -3.482 (0.072) & 0.099 \\
\enddata
\tablecomments{The magnitude errors were set to the same value.
  The errors of the fitted parameters were determined with the bootstrapping method.
}
\end{deluxetable}

\begin{deluxetable}{cccccccc}
\tabletypesize{\scriptsize}
\tablecaption{sample II PLR fit parameters\label{table_PLRs_sample_II}}
\tablewidth{0pt}
\tablehead{
\colhead{$\#$} & \colhead{band} & \colhead{type} & \colhead{range} & \colhead{$N_{fit}$} & \colhead{a (log P = 1)} & \colhead{slope b} & \colhead{$\sigma$}
}
\startdata
1 & H & FM & all & 522 & 18.901 (0.008) & -3.057 (0.028) & 0.145 \\
2 & H & FM & log(P) $>$ 1 & 163 & 18.901 (0.021) & -2.992 (0.087) & 0.175 \\
3 & H & FO & all & 102 & 18.271 (0.033) & -3.176 (0.061) & 0.082 \\
4 & J & FM & all & 522 & 19.371 (0.010) & -2.857 (0.036) & 0.193 \\
5 & J & FM & log(P) $>$ 1 & 163 & 19.369 (0.029) & -2.785 (0.115) & 0.236 \\
6 & J & FO & all & 102 & 18.701 (0.042) & -3.000 (0.080) & 0.106 \\
7 & I & FM & all & 559 & 19.832 (0.014) & -2.610 (0.049) & 0.270 \\
8 & I & FM & log(P) $>$ 1 & 171 & 19.791 (0.039) & -2.398 (0.162) & 0.331 \\
9 & I & FO & all & 111 & 19.097 (0.061) & -2.824 (0.118) & 0.162 \\
10 & g & FM & all & 559 & 21.248 (0.029) & -1.761 (0.103) & 0.557 \\
11 & g & FM & log(P) $>$ 1 & 171 & 21.177 (0.078) & -1.410 (0.331) & 0.688 \\
12 & g & FO & all & 111 & 20.407 (0.109) & -2.071 (0.207) & 0.317 \\
\enddata
\tablecomments{The magnitude errors were set to the same value.
  The errors of the fitted parameters were determined with the bootstrapping method.
}
\end{deluxetable}

\begin{deluxetable}{ccccccc}
\tabletypesize{\scriptsize}
\tablecaption{sample II broken slope PLR fit parameters\label{table_PLRs_broken_sample_II}}
\tablewidth{0pt}
\tablehead{
\colhead{$\#$} & \colhead{band} & \colhead{$N_{fit}$} & \colhead{$b_{\log(P)\leq1}$} & \colhead{$b_{\log(P)>1}$} &\colhead{$a_{\log(P)=1}$} & \colhead{$\sigma$}
}
\startdata	
1 & $\mathrm{W}_{\mathrm{JH}}$ &  522 & -3.438 (0.040) & -3.204 (0.054) & 18.223 (0.011) & 0.133 \\
2 & H &  522 & -3.172 (0.044) & -2.914 (0.071) & 18.872 (0.013) & 0.144 \\
3 & J &  522 & -2.981 (0.057) & -2.704 (0.094) & 19.339 (0.017) & 0.192 \\
4 & $\mathrm{W}_{\mathrm{gI}}$ &  559 & -3.438 (0.034) & -3.241 (0.048) & 18.569 (0.009) & 0.116 \\
5 & I &  559 & -2.772 (0.073) & -2.394 (0.129) & 19.790 (0.023) & 0.269 \\
6 & g &  559 & -2.011 (0.146) & -1.428 (0.264) & 21.184 (0.046) & 0.555 \\
\enddata
\tablecomments{The magnitude errors were set to the same value. The errors of the fitted parameters were determined with the bootstrapping method.}
\end{deluxetable}

\begin{deluxetable}{cccccccc}
\tabletypesize{\scriptsize}
\tablecaption{sample III Wesenheit PLR fit parameters\label{table_PLRs_W_sample_III}}
\tablewidth{0pt}
\tablehead{
\colhead{$\#$} & \colhead{band} & \colhead{type} & \colhead{range} & \colhead{$N_{fit}$} & \colhead{a (log P = 1)} & \colhead{slope b} & \colhead{$\sigma$}
}
\startdata
1 & $\mathrm{W}_{\mathrm{JH}}$ & FM & all &  418 & 18.241 (0.009) & -3.347 (0.034) & 0.175 \\
2 & $\mathrm{W}_{\mathrm{JH}}$ & FM & log(P) $>$ 1 &  135 & 18.244 (0.023) & -3.301 (0.083) & 0.195 \\
3 & $\mathrm{W}_{\mathrm{gI}}$ & FM & all &  444 & 18.612 (0.006) & -3.360 (0.023) & 0.104 \\
4 & $\mathrm{W}_{\mathrm{gI}}$ & FM & log(P) $>$ 1 &  137 & 18.606 (0.015) & -3.259 (0.067) & 0.119 \\
\enddata
\tablecomments{The magnitude errors were set to the same value.
The errors of the fitted parameters were determined with the bootstrapping method.
}
\end{deluxetable}

\begin{deluxetable}{cccccccc}
\tabletypesize{\scriptsize}
\tablecaption{sample III PLR fit parameters\label{table_PLRs_sample_III}}
\tablewidth{0pt}
\tablehead{
\colhead{$\#$} & \colhead{band} & \colhead{type} & \colhead{range} & \colhead{$N_{fit}$} & \colhead{a (log P = 1)} & \colhead{slope b} & \colhead{$\sigma$}
}
\startdata
1 & H & FM & all & 418 & 18.877 (0.008) & -3.102 (0.030) & 0.140 \\
2 & H & FM & log(P) $>$ 1 & 135 & 18.879 (0.022) & -3.064 (0.090) & 0.166 \\
3 & J & FM & all & 418 & 19.336 (0.009) & -2.926 (0.034) & 0.158 \\
4 & J & FM & log(P) $>$ 1 & 135 & 19.336 (0.025) & -2.892 (0.106) & 0.185 \\
5 & I & FM & all & 444 & 19.802 (0.014) & -2.707 (0.044) & 0.236 \\
6 & I & FM & log(P) $>$ 1 & 137 & 19.817 (0.037) & -2.710 (0.146) & 0.269 \\
7 & g & FM & all & 444 & 21.160 (0.028) & -1.962 (0.088) & 0.496 \\
8 & g & FM & log(P) $>$ 1 & 137 & 21.200 (0.075) & -2.082 (0.291) & 0.561 \\
\enddata
\tablecomments{The magnitude errors were set to the same value.
The errors of the fitted parameters were determined with the bootstrapping method.
}
\end{deluxetable}

\begin{deluxetable}{ccccccc}
\tabletypesize{\scriptsize}
\tablecaption{sample III broken slope PLR fit parameters\label{table_PLRs_broken_sample_III}}
\tablewidth{0pt}
\tablehead{
\colhead{$\#$} & \colhead{band} & \colhead{$N_{fit}$} & \colhead{$b_{\log(P)\leq1}$} & \colhead{$b_{\log(P)>1}$} &\colhead{$a_{\log(P)=1}$} & \colhead{$\sigma$}
}
\startdata
1 & $\mathrm{W}_{\mathrm{JH}}$ &  418 & -3.438 (0.056) & -3.234 (0.073) & 18.219 (0.015) & 0.175 \\
2 & H &  418 & -3.176 (0.046) & -3.011 (0.071) & 18.859 (0.013) & 0.140 \\
3 & J &  418 & -2.986 (0.055) & -2.851 (0.085) & 19.321 (0.015) & 0.158 \\
4 & $\mathrm{W}_{\mathrm{gI}}$ &  444 & -3.476 (0.034) & -3.191 (0.057) & 18.583 (0.009) & 0.102 \\
5 & I &  444 & -2.770 (0.073) & -2.617 (0.118) & 19.786 (0.022) & 0.236 \\
6 & g &  444 & -1.963 (0.150) & -1.961 (0.234) & 21.160 (0.046) & 0.497 \\
\enddata
\tablecomments{The magnitude errors were set to the same value. The errors of the fitted parameters were determined with the bootstrapping method.
}
\end{deluxetable}

\section{Conclusion \label{conclusion}}

We use the PAndromeda M31 Cepheid sample \citep{K18a}, which is the largest Cepheid sample in M31 with 2639 Cepheids and combine it with the PHAT data \citep{2012ApJS..200...18D} to obtain the largest Cepheid sample in M31 in four HST bands: F160W, F110W, F814W and F475W (abbreviated as H, J, I and g). Using the outlier rejection we developed in \citet{K15} we obtain Period-Luminosity relations (PLRs) with even smaller dispersion than before while increasing the sample size. 

We analyze three different Cepheid samples. Sample I is based on the \citet{K18a} sample, while sample II additionally includes literature Cepheids. Sample III is a mean magnitude corrected sample I. Sample I consists of 418 fundamental mode (FM) Cepheids and 76 first overtone (FO) Cepheids with J and H band data and 453 FM Cepheids and 84 FO Cepheids with I and g band data. Since it is based on the K18a sample, it is very homogeneous in the way the Cepheids were identified and classified. Sample II includes 522 FM Cepheids and 102 FO Cepheids with J and H band data and 559 FM Cepheids and 111 FO Cepheids with I and g band data. This sample is our largest sample and while the underlying literature samples all select the Cepheids differently and have variously accurate periods our outlier rejection negates the inherently larger scatter the sample would have by combing different samples.  

All samples show no evidence of a broken slope. Previously in \citet{K15} we found a significant broken slope. We find that the broken slope is caused by a subsample of Cepheids that have a different color from the other Cepheids and have a high extinction but are otherwise the same. The extinction however is not the cause for the broken slope, but rather how many of this different type of Cepheids are within the sample. In \citet{K15} the number was significant enough to cause the broken slope. In this work however the increased sample size hides the influence of this Cepheid type. The consequence of this finding is that PLRs in different galaxies can have different slopes. Even if the same color cut is applied, a distinct distribution of colors could cause a different slope or even a broken slope. 

The mean magnitude corrected sample III improves the dispersion in some of the bands, while the dispersion in the H band PLR and the associated $\mathrm{W}_{\mathrm{JH}}$ PLR gets larger. The mean magnitude correction could be improved with additional data, but the approach used here to correct the random phase shows that there is no systematic effect introduced by not applying a correction. The only effect we see is that mean magnitude correction produces a consistently shallower slope, but that slope is within the error of the slope determined without correction.

Same as in K15 we find a sample selection bias on the Hubble constant. We compare our samples I and II with the R12 sample and infer from the offset how 
$\mathrm{H}_0$ would change if M31 is used as the distance anchor. We find that sample I increases $\mathrm{H}_0$ by 2.6\% and sample II by 2.4\%. This is in the order of the error budget given in \citet{2016ApJ...826...56R}. For the comparison we have to assume the slope does not change. The slope of the PLR will however change between the anchor galaxy and the target galaxies due to the selection effect that causes the broken slope. We plan to study this sample selection effect in more detail, e.g. how the color cut influences this sample selection effect.

The three different samples will be published in electronic form on the CDS.

\acknowledgments

This research was supported by the DFG cluster of excellence Origin
and Structure of the Universe’ (www.universe-cluster.de).

This research was supported by the Munich Institute for Astro- and
Particle Physics (MIAPP) of the DFG cluster of excellence "'Origin and
Structure of the Universe'".  

Some of the data presented in this paper were obtained from the Mikulski Archive for Space Telescopes (MAST). STScI is operated by the Association of Universities for Research in Astronomy, Inc., under NASA contract NAS5-26555. Support for MAST for non-HST data is provided by the NASA Office of Space Science via grant NNX09AF08G and by other grants and contracts.



\begin{thebibliography}{}
	
\bibitem[Becker et al.(2015)]{2015arXiv150707523B} Becker, M.~R., Desmond, H., Rozo, E., Marshall, P., \& Rykoff, E.~S.\ 2015, arXiv:1507.07523

\bibitem[Bhardwaj et al.(2016)]{2016AJ....151...88B} Bhardwaj, A., Kanbur, S.~M., Macri, L.~M., et al.\ 2016, \aj, 151, 88 

\bibitem[Bonanos et al.(2003)]{2003AJ....126..175B} Bonanos, A.~Z., Stanek, K.~Z., Sasselov, D.~D., et al.\ 2003, \aj, 126, 175
 
\bibitem[Bono et al.(2010)]{2010ApJ...715..277B} Bono, G., Caputo, F., Marconi, M., \& Musella, I.\ 2010, \apj, 715, 277 

\bibitem[Casertano et al.(2016)]{2016ApJ...825...11C} Casertano, S., Riess, A.~G., Anderson, J., et al.\ 2016, \apj, 825, 11 

\bibitem[Chambers et al.(2016)]{2016arXiv161205560C} Chambers, K.~C., Magnier, E.~A., Metcalfe, N., et al.\ 2016, arXiv:1612.05560

\bibitem[Clementini et al.(2017)]{2017EPJWC.15202003C} Clementini, G., Eyer, L., Muraveva, T., et al.\ 2017, European Physical Journal Web of Conferences, 152, 02003 

\bibitem[Dalcanton et al.(2012)]{2012ApJS..200...18D} Dalcanton, J.~J., Williams, B.~F., Lang, D., et al.\ 2012, \apjs, 200, 18 

\bibitem[Dolphin(2000)]{2000PASP..112.1383D} Dolphin, A.~E.\ 2000, \pasp, 112, 1383 

\bibitem[Efstathiou(2014)]{2014MNRAS.440.1138E} Efstathiou, G.\ 2014, \mnras, 440, 1138 

\bibitem[Fliri et al.(2006)]{2006A&A...445..423F} Fliri, J., Riffeser, A., Seitz, S., \& Bender, R.\ 2006, \aap, 445, 423

\bibitem[Freedman \& Madore(2010)]{2010ARA&A..48..673F} Freedman, W.~L., \& Madore, B.~F.\ 2010, \araa, 48, 673 

\bibitem[Freedman \& Madore(2011)]{2011ApJ...734...46F} Freedman, W.~L., \& Madore, B.~F.\ 2011, \apj, 734, 46 

\bibitem[Freedman et al.(2012)]{2012ApJ...758...24F} Freedman, W.~L., Madore, B.~F., Scowcroft, V., et al.\ 2012, \apj, 758, 24 

\bibitem[Garc{\'{\i}}a-Varela et al.(2013)]{2013MNRAS.431.2278G} Garc{\'{\i}}a-Varela, A., Sabogal, B.~E., \& Ram{\'{\i}}rez-Tannus, M.~C.\ 2013, \mnras, 431, 2278 

\bibitem[Gieren et al.(2015)]{2015ApJ...815...28G} Gieren, W., Pilecki, B., Pietrzy{\'n}ski, G., et al.\ 2015, \apj, 815, 28 

\bibitem[Hoffmann \& Macri(2015)]{2015AJ....149..183H} Hoffmann, S.~L., \& Macri, L.~M.\ 2015, \aj, 149, 183 

\bibitem[Hoffmann et al.(2016)]{2016ApJ...830...10H} Hoffmann, S.~L., Macri, L.~M., Riess, A.~G., et al.\ 2016, \apj, 830, 10 

\bibitem[Hubble(1929)]{1929PNAS...15..168H} Hubble, E.\ 1929, Proceedings of the National Academy of Science, 15, 168

\bibitem[Humphreys et al.(2013)]{2013ApJ...775...13H} Humphreys, E.~M.~L., Reid, M.~J., Moran, J.~M., Greenhill, L.~J., \& Argon, A.~L.\ 2013, \apj, 775, 13

\bibitem[Inno et al.(2013)]{2013ApJ...764...84I} Inno, L., Matsunaga, N., Bono, G., et al.\ 2013, \apj, 764, 84 

\bibitem[Inno et al.(2015)]{2015A&A...576A..30I} Inno, L., Matsunaga, N., Romaniello, M., et al.\ 2015, \aap, 576, A30 

\bibitem[Inno et al.(2016)]{2016ApJ...832..176I} Inno, L., Bono, G., Matsunaga, N., et al.\ 2016, \apj, 832, 1

\bibitem[Kaluzny et al.(1998)]{1998AJ....115.1016K} Kaluzny, J., Stanek, K.~Z., Krockenberger, M., et al.\ 1998, \aj, 115, 1016 

\bibitem[Kaluzny et al.(1999)]{1999AJ....118..346K} Kaluzny, J., Mochejska, B.~J., Stanek, K.~Z., et al.\ 1999, \aj, 118, 346 

\bibitem[Kodric et al.(2013)]{K13} Kodric, M., Riffeser, A., Hopp, U., et al.\ 2013, \aj, 145, 106 (K13)

\bibitem[Kodric et al.(2015)]{K15} Kodric, M., Riffeser, A., Seitz, S., et al.\ 2015, \apj, 799, 144 (K15)

\bibitem[Kodric et al.(2018a)]{K18a} Kodric, M., Riffeser, A., Hopp, U., et al.\ 2018, arXiv:1806.07895 (K18a)

\bibitem[Leavitt(1908)]{1908AnHar..60...87L} Leavitt, H.~S.\ 1908, Annals of Harvard College Observatory, 60, 87 

\bibitem[Leavitt \& Pickering(1912)]{1912HarCi.173....1L} Leavitt, H.~S., \& Pickering, E.~C.\ 1912, Harvard College Observatory Circular, 173, 1 

\bibitem[Lee et al.(2012)]{2012AJ....143...89L} Lee, C.-H., Riffeser, A., Koppenhoefer, J., et al.\ 2012, \aj, 143, 89

\bibitem[Lee(2017)]{2017arXiv170102507L} Lee, C.-H.\ 2017, arXiv:1701.02507

\bibitem[Madore \& Freedman(1991)]{1991PASP..103..933M} Madore, B.~F., \& Freedman, W.~L.\ 1991, \pasp, 103, 933

\bibitem[Majaess et al.(2011)]{2011ApJ...741L..36M} Majaess, D., Turner, D., \& Gieren, W.\ 2011, \apjl, 741, L36 

\bibitem[Mochejska et al.(1999)]{1999AJ....118.2211M} Mochejska, B.~J., Kaluzny, J., Stanek, K.~Z., Krockenberger, M., \& Sasselov, D.~D.\ 1999, \aj, 118, 2211 

\bibitem[Montalto et al.(2009)]{2009A&A...507..283M} Montalto, M., Seitz, S., Riffeser, A., et al.\ 2009, \aap, 507, 283 

\bibitem[Ngeow et al.(2008)]{2008A&A...477..621N} Ngeow, C., Kanbur, S.~M., \& Nanthakumar, A.\ 2008, \aap, 477, 621

\bibitem[Riess et al.(2012)]{2012ApJ...745..156R} Riess, A.~G., Fliri, J., \& Valls-Gabaud, D.\ 2012, \apj, 745, 156 (R12)

\bibitem[Riess et al.(2016)]{2016ApJ...826...56R} Riess, A.~G., Macri, L.~M., Hoffmann, S.~L., et al.\ 2016, \apj, 826, 56 

\bibitem[Sandage et al.(2009)]{2009A&A...493..471S} Sandage, A., Tammann, G.~A., \& Reindl, B.\ 2009, \aap, 493, 471

\bibitem[Schlafly \& Finkbeiner(2011)]{2011ApJ...737..103S} Schlafly, E.~F., \& Finkbeiner, D.~P.\ 2011, \apj, 737, 103 

\bibitem[Schlegel et al.(1998)]{1998ApJ...500..525S} Schlegel, D.~J., Finkbeiner, D.~P., \& Davis, M.\ 1998, \apj, 500, 525 

\bibitem[Soszy{\'n}ski et al.(2005)]{2005PASP..117..823S} Soszy{\'n}ski, I., Gieren, W., \& Pietrzy{\'n}ski, G.\ 2005, \pasp, 117, 823 

\bibitem[Soszy{\'n}ski et al.(2015)]{2015AcA....65..297S} Soszy{\'n}ski, I., Udalski, A., Szyma{\'n}ski, M.~K., et al.\ 2015, \actaa, 65, 297 

\bibitem[Stanek et al.(1998)]{1998AJ....115.1894S} Stanek, K.~Z., Kaluzny, J., Krockenberger, M., et al.\ 1998, \aj, 115, 1894 

\bibitem[Stanek et al.(1999)]{1999AJ....117.2810S} Stanek, K.~Z., Kaluzny, J., Krockenberger, M., et al.\ 1999, \aj, 117, 2810 

\bibitem[Udalski et al.(2015)]{2015AcA....65....1U} Udalski, A., Szyma{\'n}ski, M.~K., \& Szyma{\'n}ski, G.\ 2015, \actaa, 65, 1 

\bibitem[Vilardell et al.(2006)]{2006A&A...459..321V} Vilardell, F., Ribas, I., \& Jordi, C.\ 2006, \aap, 459, 32

\bibitem[Wagner-Kaiser et al.(2015)]{2015MNRAS.451..724W} Wagner-Kaiser, R., Sarajedini, A., Dalcanton, J.~J., Williams, B.~F., \& Dolphin, A.\ 2015, \mnras, 451, 724 



\end{thebibliography}
\end{document}